\documentclass[12pt,preprint]{aastex}

\slugcomment{Astrophysical Journal, in press}

\shorttitle{CHANDRA OBSERVATION OF NGC~4649}
\shortauthors{RANDALL, SARAZIN, \& IRWIN}

\begin{document}

\title{\textit{Chandra} Observation of Diffuse Gas and LMXBs in the
Elliptical Galaxy NGC 4649 (M60)}

\author{Scott W. Randall\altaffilmark{1},
Craig L. Sarazin\altaffilmark{1}, and
Jimmy A. Irwin\altaffilmark{2,3}}

\altaffiltext{1}{Department of Astronomy, University of Virginia,
P. O. Box 3818, Charlottesville, VA 22903-0818;
swr3p@virginia.edu, sarazin@virginia.edu}

\altaffiltext{2}{Department of Astronomy, University of Michigan,
Ann Arbor, MI 48109-1090; jirwin@astro.lsa.umich.edu}

\altaffiltext{3}{Chandra Fellow}

\begin{abstract}
We present a {\it Chandra} X-ray observation of the X-ray
bright E2 elliptical galaxy NGC~4649.  In addition to bright diffuse
emission, we resolve 165 discrete sources, most of which are
presumably low-mass X-ray binaries (LMXBs).
As found in previous studies, the luminosity
function of the resolved sources is well-fit by a broken power-law.
In NGC~4697 and NGC~1553, the break luminosity was comparable
to the Eddington luminosity of a $1.4\,M_{\odot}$ neutron star.
One possible interpretation of this result is that those sources with
luminosities above the break are accreting black holes and those below
are mainly accreting neutron stars.
The total X-ray spectrum of the resolved sources
is well-fit by a hard power-law, while the diffuse spectrum requires a
hard and a soft component, presumably due to the relatively soft diffuse gas
and the harder unresolved sources.  We also find evidence for
structure in the diffuse emission near the center of NGC~4649.
Specifically, there appear to be bright ``fingers'' of emission
extending from the center of the galaxy and a 5\arcsec\ long bar at
the center of the galaxy.
The fingers are morphologically similar to radial features seen in
two-dimensional hydrodynamic simulations of cooling flows in
elliptical galaxies, and although their other properties do not match the
predictions of the particular simulations used we conclude that the
radial fingers might be due to convective motions of hot outflowing
gas and cooler inflowing gas.
The bar is coincident with the central
extended radio source; we conclude that the bar may be caused by weak
shocks in the diffuse gas from
an undetected low-luminosity
active galactic nucleus (AGN).
\end{abstract}


\section{Introduction} \label{sec:intro}

By now it has been well established that the X-ray emission from
early-type galaxies has several components.  These galaxies can
roughly be grouped into two categories based on the ratio of the X-ray
to optical luminosity $L_X/L_B$: the X-ray bright galaxies which have
large values of $L_X/L_B$, and the X-ray faint galaxies which have
small values of $L_X/L_B$.
In the X-ray bright elliptical and S0 galaxies, the X-ray emission is
dominated by thermal emission from interstellar gas
at a temperature of $kT \approx 1$ keV.
The spectra
of their X-ray faint counterparts tend to require two components:
a soft thermal component with $kT \approx 0.3$ keV,
and a hard component which has been fit as either thermal
bremsstrahlung with $kT \ga 5$ keV or a power-law
(Fabbiano, Kim, \& Trinchieri 1994; Matsumoto et al.\ 1997; Allen, di
Matteo, \& Fabian 2000; Blanton, Sarazin, \& Irwin 2001).
The luminosity of the hard component varies roughly
proportional to the optical luminosity, suggesting that the origin of
this component is low-mass X-ray binaries (LMXBs) similar to those
seen in our own Galaxy (Trinchieri \& Fabbiano 1985).  The high
spatial resolution of the {\it Chandra} X-ray Observatory has allowed
much of the hard component to be resolved into individual sources,
thereby demonstrating that this hard component is indeed from
individual sources
(e.g.,
Sarazin, Irwin, \& Bregman 2000, 2001; Blanton et al.\ 2001).

In this paper, we present the results of a {\it Chandra} observation
of the X-ray bright elliptical galaxy NGC~4649 (M60).
This is an E2 elliptical galaxy in the Virgo cluster.
NGC~4649 has a close companion galaxy NGC~4647, which
is an Sc galaxy.
This pair of galaxies is also referred to as Arp~116 or VV~206.
With a third more distant galaxy, this pair forms a group of galaxies
WBL~421
(White et al.\ 1999).
NGC~4647 was the host of the Type-I supernova SN1979a
(Barbon et al.\ 1984).

We adopt a distance for NGC~4649 of 16.8 Mpc,
based on the method of surface brightness fluctuations
(Tonry et al.\ 2001).
This is consistent with the corrected recession velocity distance in
Faber et al.\ (1989)
if the Hubble constant is 79 km s$^{-1}$ Mpc$^{-1}$.
Unless otherwise noted, all the uncertainties quoted are at the 90\%
confidence level.

\section{Observation and Data Reduction} \label{sec:obs}

NGC~4649 was observed on 2000 April 20 on the ACIS-S3 CCD
operated at a temperature of -120 C and with a frame time of 3.2 s.
In addition to the S3 chip, the ACIS chips
I2, I3, S1, S2 and S4
were also turned
on for the duration of the observation.
The pointing was determined so that the entire galaxy was located on the S3
chip and so that the center of the galaxy was not on a node boundary of the
chip.
Although a number of serendipitous sources are seen on the other chips,
the analysis of NGC 4649 in this paper will be based on data from the
S3 chip alone.
The data were telemetered in Faint mode, and
only events with ASCA grades of 0,2,3,4, and 6 were included.
We excluded bad pixels, bad columns, and the columns next to
bad columns and to the chip node boundaries.
We checked for periods of incorrect aspect solution, and none were found.
The total exposure for the S3 chip was 36,780 s.

\begin{figure}
\plotone{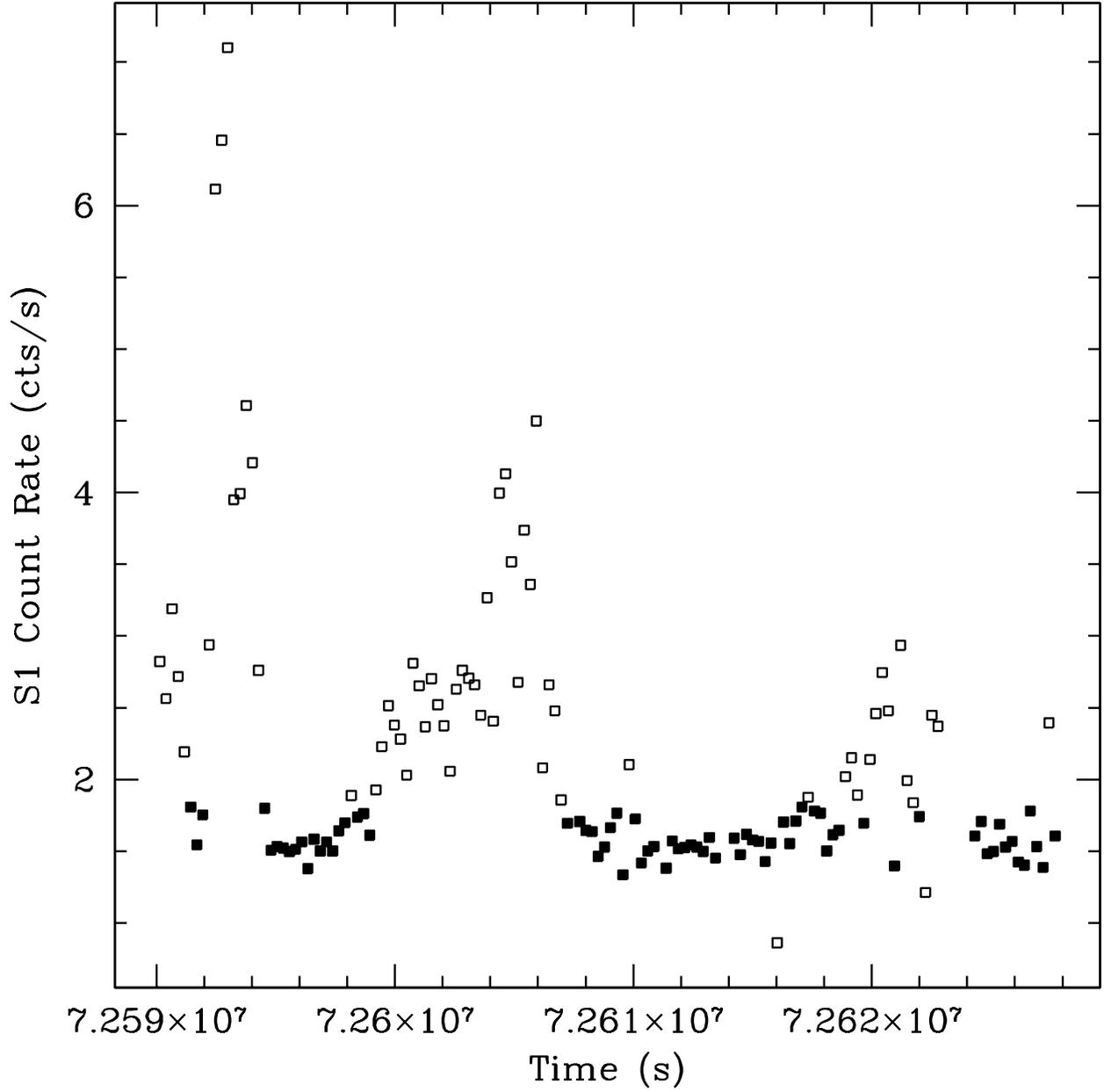}
\caption{Light curve of the entire S1 chip in the 0.3-10.0 keV band.
The open squares mark time periods which were not included in the cleaned data.
The average rate after removal of background flares was 1.61 cnt s$^{-1}$.
Time bins are 256 s.
\label{fig:s1_lcurve}}
\end{figure}

{\it Chandra} is known to encounter periods of high background
(``background flares''), which especially affect the
backside-illuminated S1 and S3
chips\footnote{See \url{http://hea-www.harvard.edu/$\sim$maxim/axaf/acisbg/}.}.
We determined the background count rate, using the S1 chip to avoid the
enhanced flux due to the galaxy NGC~4649 on the S3 chip.
Unfortunately, much of the exposure was affected by background flares;
the light curve of the total count rate (0.3--10 keV) for the S1 chip is
shown in Figure~\ref{fig:s1_lcurve}.
We used the program {\sc lc\_clean}\footnotemark[4]
written by Maxim Markevitch to remove periods of high background and
data drop-outs.
The cleaned light curve for the S1 chip is shown as the filled squares in
Figure~\ref{fig:s1_lcurve}.
After cleaning, 19,303 of the total 36,780 s of exposure time remained.

The expected quiescent rate for the S1 chip during this period
in the 0.3-10.0 keV band is about\footnotemark[4] 1.41 cnt s$^{-1}$.
After the removal of background flares, the mean quiescent S1 rate during our
observation was 1.61 cnt s$^{-1}$.
The other, front-side illuminated chips also show excess background.
By examining different energy bands we determined that most of the
excess background in the S1 chip was below 5 keV, with our quiescent rates
being in excess of the expected rates by about 21\% in the 0.3-1.0 keV band,
27\% in the 1.0-5.0 keV band, and 9\% in the 5.0-10.0 keV band.
We note that NGC~4649 is located both at the outer edge of the Virgo cluster
and at the edge of the Galactic North Polar Spur,
both of which may contribute to the soft X-ray background near NGC~4649
(B\"ohringer et al.\ 1994).
We examined
{\it ROSAT} observations of the same region of the sky covered by the S1
chip and determined a
background count rate of 0.031 $\pm$ 0.01 cnt s$^{-1}$ in the {\it
ROSAT} PSPC R4-R7 band.
This roughly translates to 0.115 cnt s$^{-1}$
for the {\it Chandra} S1 chip in the 0.3--10 keV energy band, thereby
accounting for about 60\% of the excess background emission we
detected with the S1 chip.
Here, {\sc pimms} was used to convert the {\it ROSAT} count rate into
a {\it Chandra} 0.3--10 keV count rate by assuming a Raymond-Smith (RS)
spectral model with temperature $kT = 0.862$ keV and solar abundance.
This model was chosen as a compromise between the spectrum of the North
Polar Spur (NPS) (Snowden et al.\ 1997) and that of diffuse emission from the Virgo
cluster (B\"ohringer et al.\ 1994).
The excess background at hard energies (9\%) is probably due to
a slightly elevated particle background, even when the background flares
are removed.
If we assume that this is the case and renormalize the blank sky data
used to determine the quiescent rate given above so that the count
rate in the 5--10 keV band matches what we observed with the S1 chip
(see \S~\ref{sec:spectra}),
then the excess emission from the Virgo cluster and the NPS determined
from the ROSAT observations can
completely account for the remaining excess background emission in the
0.3--5 keV energy band.  We therefore conclude that the excess
background emission we detect is a combination of emission from the
Virgo cluster, the NPS, and an elevated particle background.

Three of the X-ray point sources detected on the S3 image have optical
counterparts with positions listed in the
the U.S. Naval Observatory (USNO)
A2.0 optical catalog
(Monet et al.\ 1998).
The optical and X-ray positions all agree to better than 2\arcsec.
Thus, we believe that the absolute positions derived from the X-ray
images are accurate to better than 2\arcsec.

\section{X-ray Image} \label{sec:image}

%
%

The raw {\it Chandra} S3 chip X-ray image is shown in
Figure~\ref{fig:xray_whole} for the cleaned 19 ksec exposure in the
0.3--10 keV energy band.  The sky background has not been subtracted
from this image, nor has the exposure map been applied.
Many discrete sources are evident.
Also visible is bright diffuse emission at the center of the galaxy,
which swamps emission from discrete sources in that region.
%
%
In order to image the fainter, more diffuse emission, we adaptively
smoothed the cleaned 19 ksec exposure {\it Chandra} S3 X-ray image
to a minimum signal-to-noise ratio of 3 per smoothing beam.  The
resulting image is shown in Figure~\ref{fig:xray_smo}.
The image was corrected for exposure and background.
This image shows rather extended diffuse emission from NGC~4649 as
well as the point sources.
In addition to some extended features in the outer parts of the
galaxy,
the image appears to show faint radial ``fingers'' of emission reaching
out from the center of NGC~4649, which are discussed further in
\S~\ref{sec:radial_features}.

\section{Resolved Sources} \label{sec:sources}

\subsection{Detections} \label{sec:src_detect}

\begin{figure}
\plotone{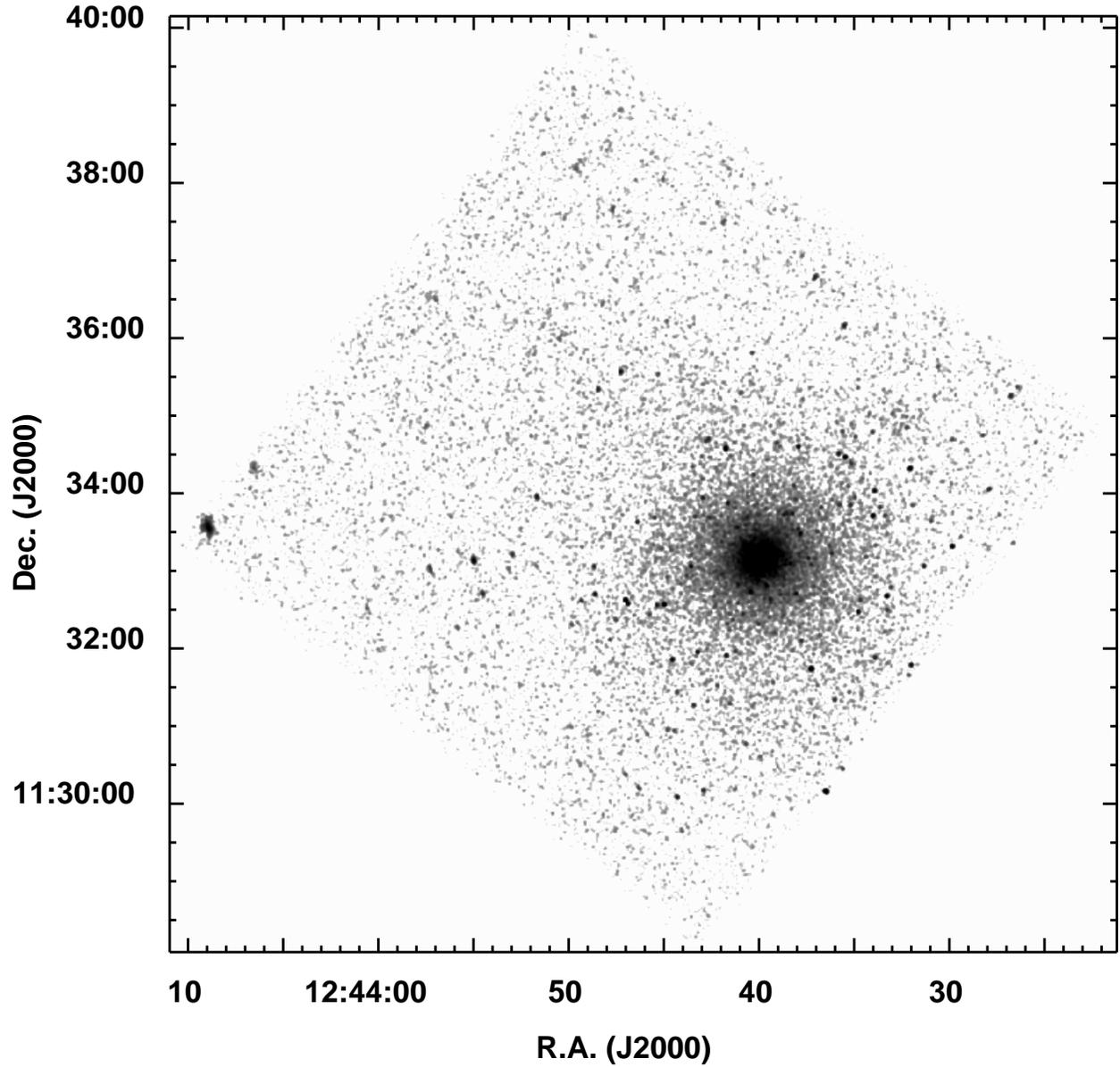}
\caption{Raw {\it Chandra} S3 image of NGC~4649, cleaned of background
flares but uncorrected for
background or exposure.
The image was smoothed with a 2 pixel gaussian to make the point sources
easier to see.
The greyscale is logarithmic and ranges from about 0.06 to
4 cnt pix$^{-1}$.
Both discrete sources and diffuse emission are visible.
\label{fig:xray_whole}}
\end{figure}

\begin{figure}
\plotone{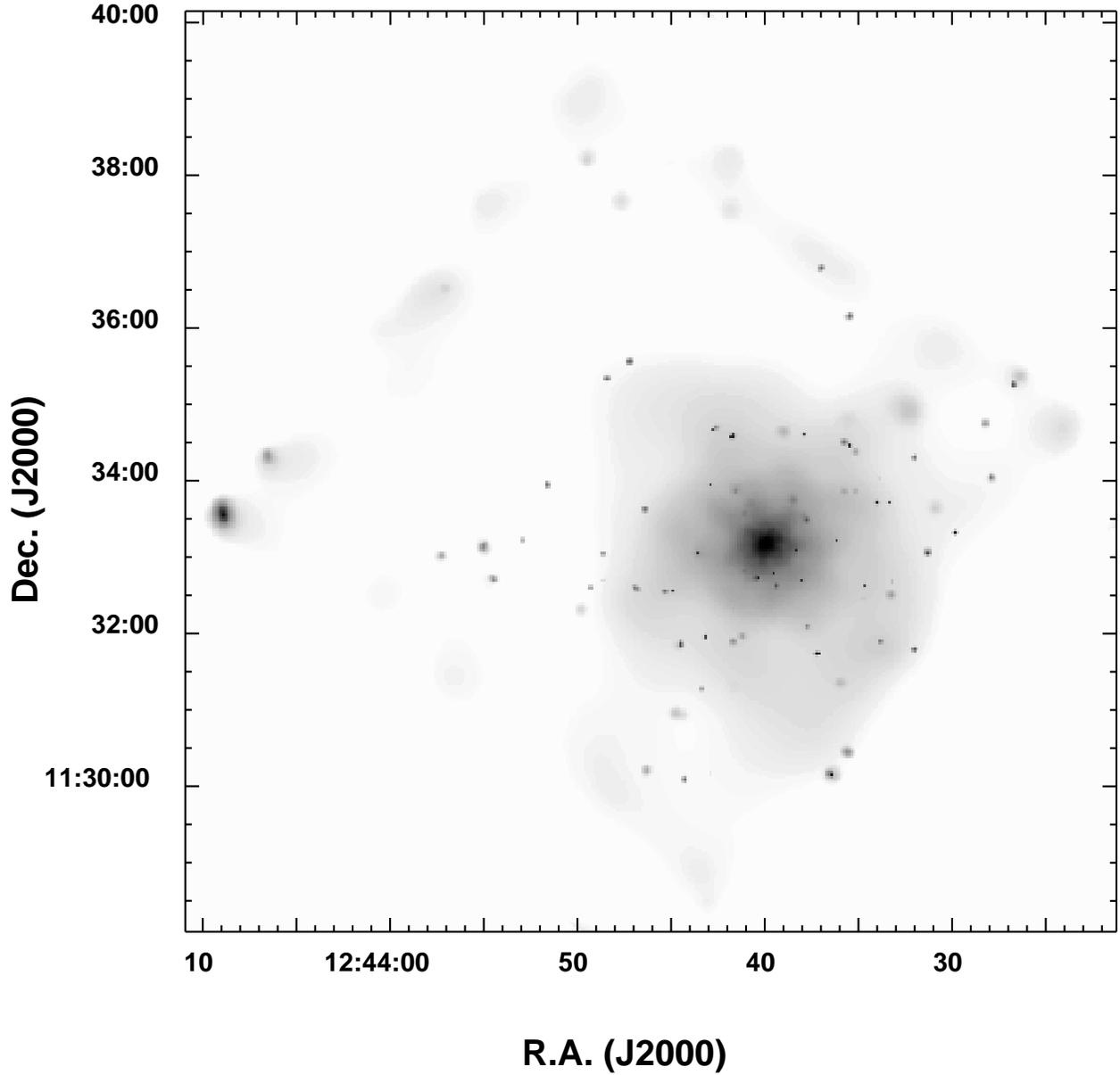}
\caption{
Adaptively smoothed {\it Chandra} S3 image of NGC~4649,
cleaned of background flares and corrected for exposure and background.
The greyscale is logarithmic and ranges from $1 \times 10^{-5}$
to $5 \times 10^{-3}$ cnt pix$^{-1}$ s$^{-1}$.
Some of the extended sources near the edge of the field may be artifacts
due to the large exposure correction in this region.
\label{fig:xray_smo}}
\end{figure}

The discrete X-ray source population on the ACIS S3 image
(Figure~\ref{fig:xray_whole})
was determined using a wavelet detection algorithm
in the 0.3--10.0 keV band, and
they were confirmed with a local cell detection method.
We used the {\sc ciao}\footnote{See \url{http://asc.harvard.edu/ciao/}.}
{\sc wavdetect}
and {\sc celldetect} programs.
The high spatial resolution of {\it Chandra} implies that
the sensitivity to point sources is not affected very strongly by the
background.
Thus, the source detection was done using the entire exposure of 36,780 s,
including periods with background flares.
We also did the source detection on the data after the removal of the
background flares;
the stronger sources were all found in both cases, but the very faintest
sources were not detected as significantly in the cleaned exposure at
a level consistent with the much shorter exposure (19,303 vs.\ 36,780 s).
The wavelet source detection significance threshold was set at $10^{-6}$,
which implies that $<$1 false source
(due to a statistical fluctuation in the background) would be detected
in the entire S3 image.
This significance threshold approximately corresponds to requiring
that the source flux be determined to better than 3-$\sigma$.
Fluxes were corrected for exposure and the instrument
point-spread-function (PSF) using the standard {\sc ciao} CALDB PSF
tables\footnote{See \url{http://asc.harvard.edu/caldb/}.}.
All of the source detections were verified by visually examining the
image to ensure that no obvious errors were made during the detection process.

Table~\ref{tab:src} lists the 165 discrete sources detected by this
technique,
sorted in order of increasing distance $d$ from the center of NGC~4649.
Columns 1-7  give
the source number,
the IAU name,
the source position (J2000),
the projected distance $d$ from the center of NGC~4649,
the count rate and the 1-$\sigma$ error,
and
the signal-to-noise ratio SNR for the count rate.
Since we did not detect a distinct source at the center of the galaxy,
we adopted the central position from 4.86 GHz radio observations of
R.A.\ =\ 12$^{\rm h}$43$^{\rm m}$40\fs02, and
Dec.\ =\ +11\arcdeg33\arcmin10\farcs2 (J2000; Condon et al.\ 1991).
The accuracy of this position is $\la 1\arcsec$.
Because of the uncertainty in the exact position of the center of the
galaxy,
the values of $d$ might be off by $\approx$2\arcsec.
The statistical errors in the positions of most of the sources are quite
small ($\sim 0.2\arcsec$),
and the overall absolute errors are probably $\sim$1\arcsec\ near the
center of the field, with larger errors near the outside of the
field.
These absolute errors are typical for observations during this
period\footnote{See
\url{http://asc.harvard.edu/mta/ASPECT/cal\_review\_aspect.ps}.}.

%
%

Over most of the image, the minimum detectable count rate was about
$3.1 \times 10^{-4}$ cnt s$^{-1}$
($L_X = 7 \times 10^{37}$ ergs s$^{-1}$ at the NGC~4649 distance)
in the 0.3--10 keV band.
The detection limit is slightly higher (by $\la$ 30\%) at large distances
$d$ where the PSF is larger, particularly at the northern edge of the field.
The minimum detectable flux is also much higher near the center of
NGC~4649, where the diffuse emission by interstellar gas is very
bright.
This effect is obvious in Table~\ref{tab:src}.
With the exception of Src.~1,
which is a detection of the extended peak in the diffuse emission,
no sources are seen within 12\arcsec\ of the center.
If the X-ray source distribution followed the optical light in NGC~4649,
one would have expected at least 10 sources in this region.
Within the central 30\arcsec\ in radius, all of the sources are brighter
than $\ga$$10 \times 10^{-4}$ cnt s$^{-1}$, which is $\ga$3 times the
detection limit at larger distances.
 From the values of the diffuse plus background count rates found by
the {\sc wavdetect} algorithm, we find that the source detection threshold
is increased for $d \la 70\arcsec$.

Because the diffuse galactic X-ray emission is soft
(\S~\ref{sec:diffuse_spec}) while the sources are mainly hard,
we also did the source detection on a hard band (2--10 keV) image
both for the entire exposure and the cleaned exposure
in the hope that this would allow sources to be detected more readily in
the inner parts of the galaxy.
However, we did not detect any additional sources in the hard band.

Our detection limit for sources should result in $<$1 false source
(due to a statistical fluctuation) in the entire S3 field of view.
However, some of the detected sources are likely to be unrelated foreground
or (more likely) background objects.
Based on the source counts in
Brandt et al.\ (2000) and
Mushotzky et al.\ (2000),
we would expect $\approx$10 such unrelated sources in our observation.
These should be spread out fairly uniformly over the S3 image
(Figure~\ref{fig:xray_whole}; see Giacconi et al.\ 2001), except for the reduced sensitivity at
the center of NGC~4649 due to bright diffuse emission and at the
outer edges of the field due to reduced exposure and increased PSF.
Thus, the unrelated sources should mainly be found at larger distances
from the optical center of NGC~4649 (the bottom part of
Table~\ref{tab:src}), while the sources associated with NGC~4649 should be
concentrated to the center of the galaxy.

We searched for variability in the X-ray emission of the resolved sources
over the duration of the {\it Chandra} observation using
the Kolmogoroff-Smirnov
test (see Sarazin et al.\ 2001).
In most cases, the tests were inconclusive.
For only two of the sources (Srcs.~76 \& 98) was the probability that
they were constant $\ll 1$\%;
these sources are marked with an ``h'' in the Notes column of
Table~\ref{tab:src}.
Src.~76 appeared to brighten during the observation by
a factor of $\sim$ 1.6, while Src.~98 faded so
as to be undetectable (by a factor of $\ga$ 1.5).

\subsection{Identifications} \label{sec:src_id}

As noted above, there was only one source, Src.~1, detected within
12$\arcsec$ of the center of NGC~4649.
We believe that
this detection does not represent an individual source, but rather a
structural feature in the diffuse emission.
This source is 5.22 times wider (FWHM) than the PSF of a point source
at the same location, and is wider than any other detected source in
the image.
We will therefore drop Src.~1 from further discussion of the resolved sources.

%
%

\begin{figure}
\plotone{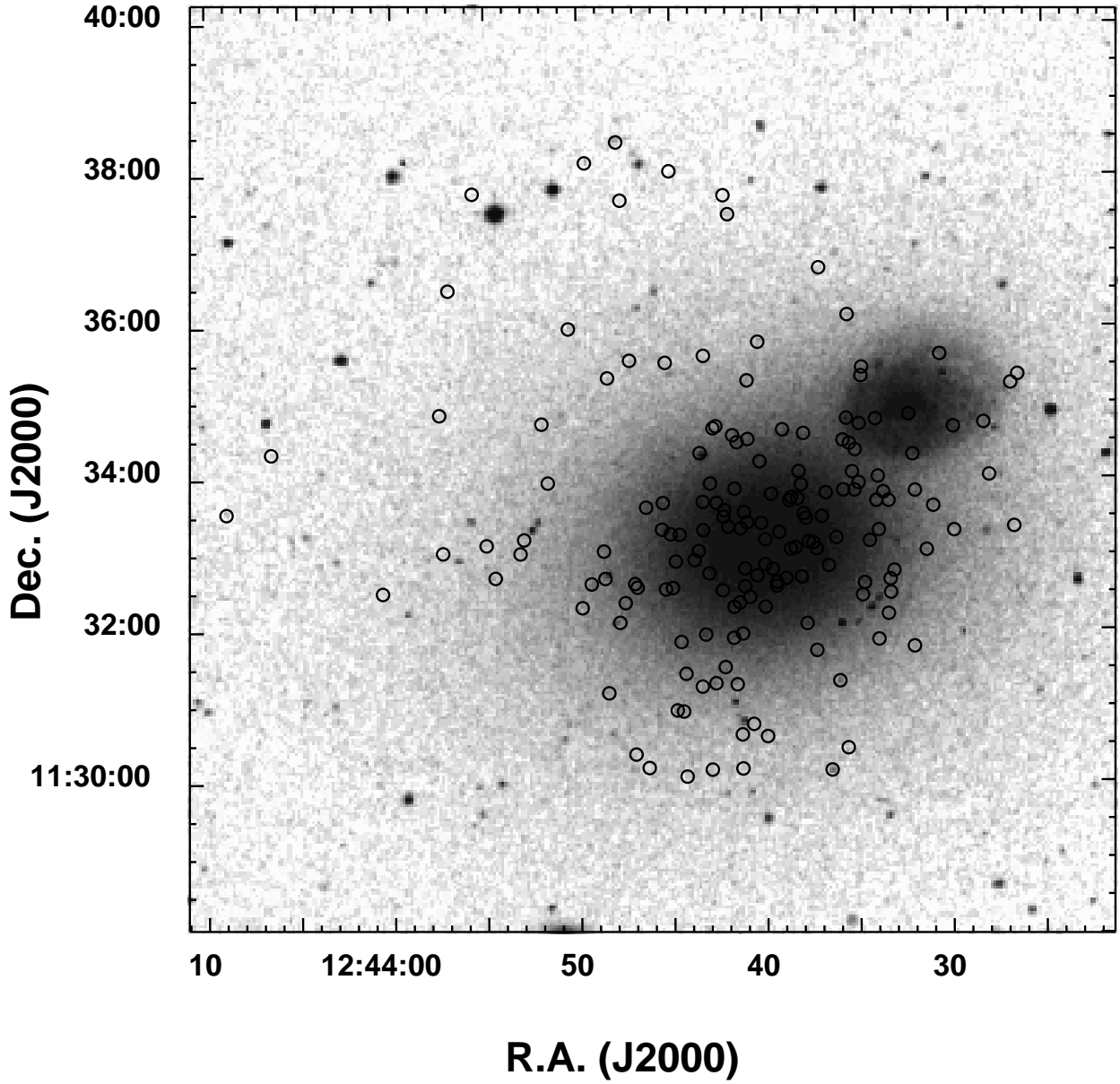}
\caption{DSS optical image of NGC~4649.
The region covered is the same as in Figures~\protect\ref{fig:xray_whole}
\& \protect\ref{fig:xray_smo}.
The circles indicate the positions of the detected X-ray sources.
\label{fig:dss_field}}
\end{figure}

We compared the positions of the X-ray sources with the
Digital Sky Survey (DSS) image of this region
(Fig.~\ref{fig:dss_field}).
Fourteen of the sources had possible faint optical counterparts on
this image.
These are all marked with note ``d'' in Table~\ref{tab:src}.
All of the
possible X-ray/optical source matches occur in the outer part of the
field away from the center of NGC~4649 since the optical detection of
sources near the center of the galaxy is difficult.
Three of the possible optical counterparts (Srcs. 136, 162, 165)
have positions listed in the USNO-A2.0 catalog
(Monet et al.\ 1998).
These sources are indicated by a ``e'' in the Notes column of
Table~\ref{tab:src}.
We find that two of the USNO optical sources have positions which match
those of X-ray sources to within 1\arcsec, and one matches to within
2\arcsec\ (Table~\ref{tab:src}).
Given the density of USNO sources in the field, one would expect about
0.3 misidentifications within 2\arcsec\ of the {\it Chandra} sources.

We also compared the positions of the X-ray sources with those of
globular clusters (GCs) in NGC~4649.
In the outer parts of the galaxy ($d \ga 3\arcmin$), a list is
provided by Hanes (1977), based on ground-based optical images.
Srcs. 107, 136, 157, 160, \& 162 correspond to Hanes clusters
36, 49, 19, 25, \& 17, respectively.
Based on the positional accuracy, the numbers of globular clusters,
the number of X-ray sources, and the area of the sky covered,
we estimate that $\sim$0.7 X-ray sources would be expected to agree
with Hanes GCs by coincidence.
It is worth noting that, at the distance to NGC~4649, globular clusters are
not resolved in ground-based optical images.
As a result, as many as half of these GCs which are located at large radii
might be unrelated faint optical objects, rather than globular clusters.
Kundu \& Whitmore (2001) have determined the globular cluster population
of a region of NGC~4649 to the northwest based on
{\it Hubble Space Telescope (HST)} images.
Confusion with background sources should not be significant in the
{\it HST} data as these images slightly resolve the globular clusters.
A. Kundu (private communication) very kindly provided the positions,
magnitudes, and colors of these globulars.
We found that 24 of these GCs were within 0\farcs9 of the positions of X-ray
sources.
Two of these may be associated with NGC~4647.
There were 445 GCs associated with NGC~4649 and 47 X-ray sources in
the region covered by the {\it HST} image
and not associated with NGC~4647; one would expect $<$3
associations to occur at random.
Thus, it is likely that most of these identifications are real.
Within the region of NGC~4649 covered by the {\it HST} image, roughly
47\% of the X-ray sources are identified with globular clusters.

A serendipitous {\it ROSAT} source RX J1244.1+1134
(Romer et al.\ 2000)
was detected in
the PSPC pointed observation of NGC~4649 at a position which is
consistent with Srcs.~164 \& 165.
 From comparison of our {\it Chandra} image with the {\it ROSAT}
PSPC image, we believe that this source is actually a blend of
these two sources, plus another source (comparable to Src.~165 in
brightness), but located just off the S3 chip on the S4 chip.
Romer et al.\ also indicated that this source is probably a blend.

As shown in Figure~\ref{fig:dss_field}, NGC~4649 has a nearby companion galaxy,
the Sc galaxy NGC~4647.
Some of the X-ray sources are projected on the optical image of this galaxy.
These sources are marked with note ``g'' in Table~\ref{tab:src}.
In particular, Src.~119 is located within 4\arcsec\ of the nucleus of
NGC~4647.
This galaxy also was the host of the Type I supernova SN1979a
(Barbon et al.\ 1984).
We were unable to find any very accurate position for the supernova in
the literature;
however, none of the X-ray sources are within 20\arcsec\ of the approximate
position given by Barbon et al.\ (1984).

\subsection{X-ray Luminosities and Luminosity Function} \label{sec:src_lum}

The count rates for the sources were converted into unabsorbed luminosities
(0.3-10 keV) assuming that all of the sources were at the distance of
NGC~4649, which we take to be 16.8 Mpc
(Tonry et al.\ 2001).
We adopted the best-fit {\it Chandra} X-ray spectrum of the resolved
sources within the inner 1 effective radius
($R_{\rm eff} = 82\arcsec$, van der Marel 1991; Table~\ref{tab:spectra} below).
The factor for converting the count rate (0.3--10 keV)
into the unabsorbed luminosity $L_X$ (0.3--10 keV) was
$2.25 \times 10^{41}$ ergs cnt$^{-1}$.
The resulting X-ray luminosities are given in column~8 of
Table~\ref{tab:src} in units of $10^{37}$ ergs s$^{-1}$,
and range roughly from $7 \times 10^{37}$ to $8 \times 10^{39}$ ergs s$^{-1}$.

%
%

\begin{figure}
\vskip4truein
\includegraphics{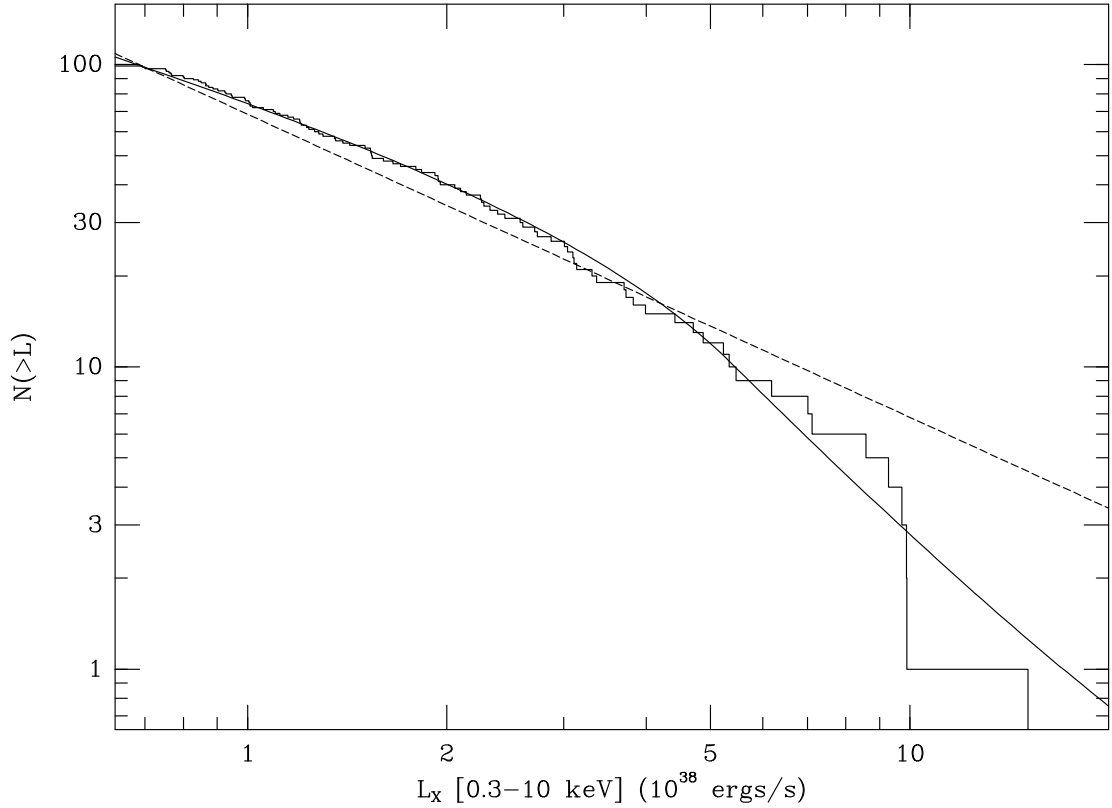}
\caption{Solid histogram is the cumulative luminosity function of the sources
in the region of the S3 chip with
$70\arcsec \le d \le 4\arcmin$, excluding a very small region near
the chip edge.
The dashed curve is the best-fit single power-law, while the solid curve
is the best-fit broken power-law
(equation~\protect\ref{eq:xlum}).
\label{fig:xlum}}
\end{figure}

We determined the luminosity function for the sources with
$70\arcsec \le d \le 4\arcmin$, excluding a very small region near
the chip edge with a low exposure.
This includes sources which may be associated with NGC~4647;
however, when we compared the number of
sources in the region of NGC~4647 to the number of sources in other
regions of similar size at the same
projected distance from NGC~4649, we found no
evidence for an over-abundance of sources near NGC~4647.
Note that removing the region of low exposure removed Src.~136,
which has the highest luminosity of any detected source.
We did not include the center of the galaxy ($d < 70\arcsec$)
because the brightness of the diffuse gaseous emission increased the
minimum detectable source flux there.
The cumulative luminosity function of the sources from this region is shown
as a histogram in Figure~\ref{fig:xlum}, under the assumption that
all of the sources are located at the distance of NGC~4649.
We fit the luminosity function, using the same techniques as we
used previously
(Sarazin et al.\ 2000, 2001;
Blanton et al.\ 2001;
Irwin, Sarazin, \& Bregman 2001).
In fitting the luminosity function, we included a correction
for unrelated sources using the deep source counts in
Brandt et al.\ (2000) and
Mushotzky et al.\ (2000).
We first tried a single power-law model for the differential luminosity
function (dashed curve in Fig.~\ref{fig:xlum}), but it did not provide
a very good fit and could be rejected at the $>$95\% level.
A broken power-law model,
\begin{equation} \label{eq:xlum}
\frac{ d N }{ d L_{38} } = N_o \, \left\{
\begin{array}{l}
( L_{38} / L_b )^{-\alpha_l} \qquad L_{38} \le L_b \\
\\
( L_{38} / L_b )^{-\alpha_h} \qquad L_{38} >  L_b \\
\end{array}
\right. \, ,
\end{equation}
gave a good fit,
where $L_{38}$ is the X-ray luminosity (0.3--10 keV) in units
of $10^{38}$ ergs s$^{-1}$.
The best-fit values and 90\% confidence uncertainties are
$N_o = 4.0^{+9.4}_{-2.7}$,
$\alpha_l = 1.56^{+0.30}_{-0.43}$,
$\alpha_h = 3.30^{+4.07}_{-0.79}$,
and a break luminosity of
$L_b = 5.3^{+4.4}_{-2.3} \times 10^{38}$ ergs s$^{-1}$
(90\% errors).
Note that the normalization $N_o$ only applies to the area of the
galaxy used to derive the luminosity function;
the total population is probably about a factor of two larger,
assuming that the X-ray sources are distributed similarly to the
optical light in the galaxy.

The low-luminosity slope $\alpha_l$ is very similar to that found in
the elliptical NGC~4697
(Sarazin et al.\ 2000, 2001) and the Sa bulge NGC~1291
(Irwin et al.\ 2001).
In fact, Irwin et al.\ (2001) argued that the slope and normalization
(relative to the optical luminosity)
of the low end of the luminosity function in early-types galaxies and
spiral bulges might be universal.
The high luminosity slope $\alpha_h$ is steeper than in  NGC~4697,
but consistent within the errors.
The break luminosity $L_b$ is larger than has been
found in other early-type galaxies (Blanton et al.\ 2001; Sarazin et
al.\ 2001; Finoguenov \& Jones 2002; Kundu et al.\ 2002),
although it is consistent with these other measurements within the
(relatively large) errors.
Sarazin et al.\ (2000) suggested that $L_B$ corresponded to the Eddington
luminosity of a 1.4 $M_\odot$ neutron star
($L_{\rm Edd,NS} \approx 2 \times 10^{38}$ ergs s$^{-1}$),
and that this value is universal.
The luminosity function in NGC~4649 may suggest that there are variations
in $L_b$ from galaxy to galaxy, and/or that $L_b$ is larger than
$L_{\rm Edd,NS}$ by a factor of $\sim$1.5--2.
Alternatively, variations in $L_b$ might be due to distance errors.
In NGC~4649, the distance we adopted would need to be too large by $\sim$25\%
to give a best-fit break luminosity of $3 \times 10^{38}$ ergs s$^{-1}$.
This is larger than the stated statistical errors
(Tonry et al.\ 2001).

\subsection{Hardness Ratios} \label{sec:src_colors}

%
%

\begin{figure}
\vskip4truein
\includegraphics{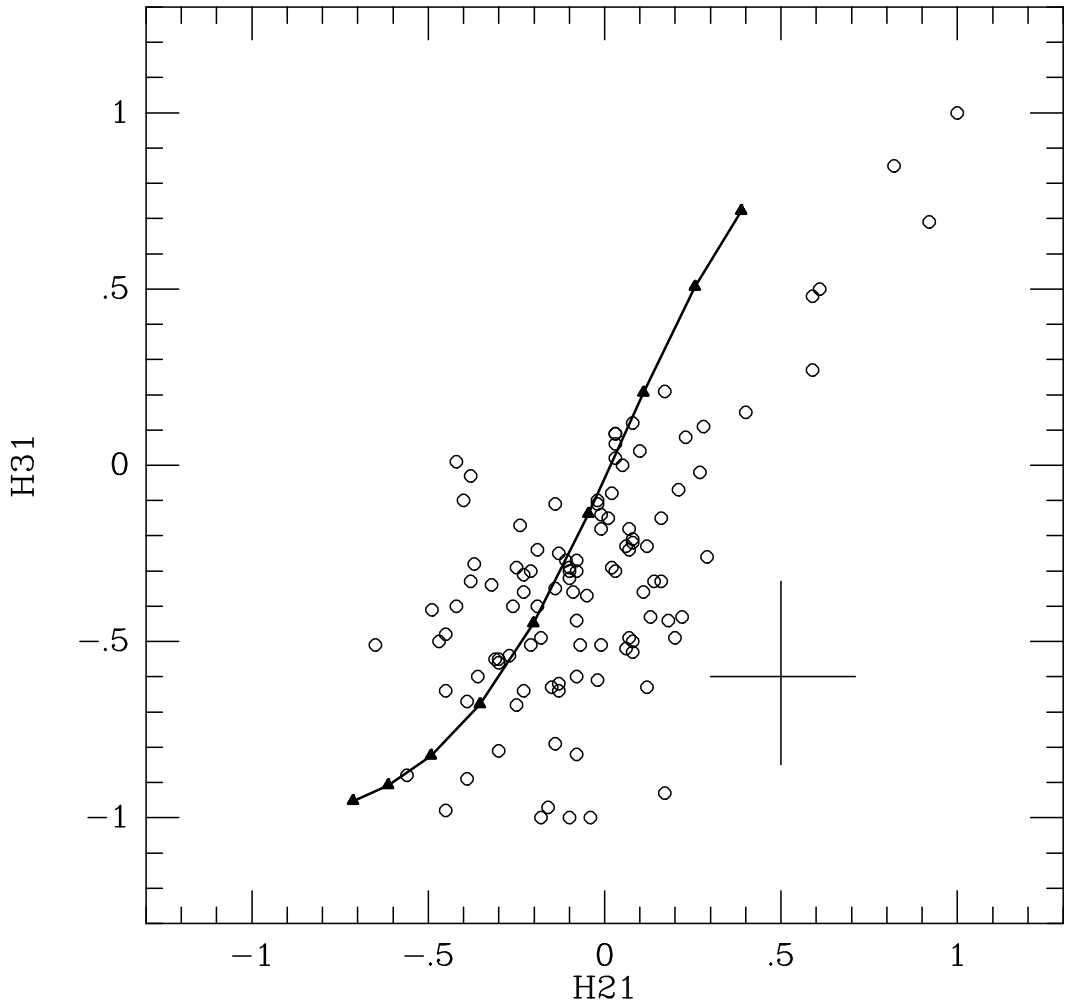}
\caption{Hardness ratios for the NGC~4649 sources with at least 20 net counts.
Here, $H21 \equiv ( M - S ) / ( M + S )$
and $H31 \equiv ( H - S ) / ( H + S ) $, where $S$, $M$, and $H$ are
the
net
counts in the soft (0.3--1 keV), medium (1--2 keV), and hard
(2--10 keV) bands, respectively.
Note that there are two sources located at $(H21,H31) = (1,1)$.
The solid line and triangles show the hardness ratios for power-law
spectral models with Galactic absorption;
the triangles indicate values of the power-law photon number index
of $\Gamma = 0$ (upper right) to 3.2 (lower left) in increments of
0.4.
The error bars at the lower right illustrate the approximate uncertainties
for a moderate flux source (one with a total of 40 counts).
\label{fig:colors}}
\end{figure}

We determined X-ray hardness ratios for the sources, using the same
techniques and definitions we used previously
(Sarazin et al.\ 2000, 2001;
Blanton et al.\ 2001;
Irwin et al.\ 2001).
Hardness ratios or X-ray colors are useful for crudely characterizing the
spectral properties of sources, and can be applied to sources which are
too faint for detailed spectral analysis.
We define two hardness ratios as H21 $\equiv ( M - S ) / ( M + S )$
and H31 $\equiv ( H - S ) / ( H + S ) $, where $S$, $M$, and $H$ are
the
net counts in the soft (0.3--1 keV), medium (1--2 keV), and hard
(2--10 keV) bands, respectively.
The hardness ratios are listed in columns 9 \& 10 of
Table~\ref{tab:src} for all of the resolved sources.
The errors in the hardnesses ratio are determined from the Poisson errors
in the original counts in the bands, and are carefully propagated so as to
avoid mathematically impossible hardness ratios;
that is, the error ranges are limited to $-1$ to 1.
Figure~\ref{fig:colors} plots H31 vs.\ H21 for all of the 107 sources with
at least 20 net counts.
For comparison, the hardness ratio for the sum of the sources is
(H21,H31) $ = (-0.09,-0.38)$.

As was also seen in NGC~4697, NGC~1553, and the bulge of NGC~1291
(Sarazin et al., 2000, 2001;
Blanton et al., 2001;
Irwin et al., 2001),
most of the sources lie along a broad diagonal swath extending roughly from
(H21,H31) $\approx (-0.5,-0.7)$ to (0.3,0.2).
For example, these colors correspond to Galactic absorption and power-law
spectra with photon indices of $\Gamma \approx -2$ to $-1$.

In Figure~\ref{fig:colors}, there are five sources with very hard spectra
(hardness ratios [H21,H31] $>$
$[0.6,0.5]$);
these are Srcs.~23, 52, 77, 84, \&
110.
These may be unrelated, strongly absorbed AGNs,
similar to the sources which produce the hard component of the
X-ray background, and which appear strongly at the faint fluxes in the
deep {\it Chandra} observations of blank fields
(Brandt et al.\ 2000;
Mushotzky et al.\ 2000;
Giacconi et al.\ 2001).
However, of these five sources, two (Srcs.~23 \& 52) have hardness
ratio errors that include the full rage of possible values and two
more (Srcs.~77 \& 84) have error ranges that include $(0,0)$.
Therefore, Src. 110 is the only source that can be said to be a
strongly absorbed background AGN with any certainty.
There also is a group of sources with hardness ratios of around
$(-0.2,-1)$; these are
Srcs.~2, 7, 10, 19, 56, 81, 129, 137, 149, 158, \& 162.
These sources have very little hard emission, and many are at large
radii.
Due to the lack of hard emission, some of these sources have very
poorly constrained H31 values (e.g. Srcs.~56, 81, 158, \& 162),
although the H21 values are known to reasonable accuracy.
Studies of other galaxies
(Sarazin et al.\ 2001)
and deep blank sky images
(e.g., Giacconi et al.\ 2001)
suggest that many of these sources may also be unrelated background sources.
However, some of these sources (particularly the ones at smaller
radii) may be sources in NGC~4649.
The total number of sources with hardness ratios of $\sim$(1,1) or
$\sim$$(-0.2,-1)$, including faint sources not plotted in
Figure~\ref{fig:colors} with large errors in the hardnesses ratios,
is $\sim$30, which is considerably higher than the $\sim$10 background
sources expected based on deep {\it Chandra} images.
Similarly, the total number of expected background sources
based on deep {\it Chandra} counts
with these hardness ratios and $\ga 20$ net
counts (e.g., those sources that would be plotted in
Figure~\ref{fig:colors})
is $\approx 6$,
as compared to the 12 sources listed above with relatively well-known
hardness ratios.
Thus, it is likely that many of these sources are associated with
NGC~4649 or NGC~4647.

\section{X-ray Spectra} \label{sec:spectra}

We used the {\sc ciao}\footnotemark[5] script {\sc psextract} to extract
spectra.
For extended or multiple sources, the response matrices were
determined using the {\sc calcrmf/calcarf}\footnote{See
\url{http://asc.harvard.edu/cont-soft/software/}.}  package written by
Alexey Vikhlinin and Jonathan McDowell, which weighted them by the X-ray
brightness in the 0.5--2 keV energy band over the corresponding image region.
We used the gain file
acisD2000-01-29gainN0003.fits
and the fef file
acisD2000-01-29fef\_piN0001.fits.
For the
individual resolved sources, local backgrounds were used.
For the diffuse spectra, we used the blank sky
background files provided by Maxim Markevitch\footnotemark[4]
to generate background spectra.
Since the count rate we detected on the S1 chip was in
excess of the quiescent background rate expected (see \S~\ref{sec:obs}),
we renormalized the blank sky background spectra to match our observed
S1 data in the 5--10 keV band.
The blank sky background spectra for the S3 chip, which was used as background for
the spectral analysis of the diffuse emission throughout, was normalized by this
same factor.
When fitting spectra we consider only the 0.7--10 keV energy band throughout
since the response
below 0.7 keV is uncertain.
Each spectrum
has been grouped to a minimum of 20 counts per pulse invariant (PI)
channel so that $\chi^2$ statistics apply.
We used {\sc xspec} to fit models to the spectra.
For all the fitted models discussed here, the
absorption column was fixed at the Galactic value
($N_H = 2.20 \times 10^{20}$ cm$^{-2}$; Dickey \& Lockman 1990).
Allowing the absorption to vary did not improve the fits significantly.

One concern with normalizing the blank sky background spectra in this way
is that the spectrum of the excess background in our observations may
differ from that in the blank sky background observations.
For example, the excess background might be due to the Virgo cluster
or the North Polar Spur.
To model this effect, we determined the spectrum of the excess background
on the S1 chip, using the normalized blank sky data as a background.
Since we expect this excess emission to be mainly due to diffuse gas
(in the Virgo cluster or North Polar Spur),
we modeled this excess emission with a mekal model.
 While we did find a statistically good fit, the temperature and abundance
were only very poorly constrained.
We included this excess background component in our fits for the diffuse
emission in NGC~4649, freezing the spectral parameters to the best fit
values obtained above.
The normalization was scaled with the area of the detector used.
Although including this component did somewhat
affect the normalizations of the other model components, the fitted
values for the temperature and abundance agreed with the results we
obtained without including this component to well within the errors.
Including this excess emission component in the fit to the outermost
annulus defined
in Table~\ref{tab:spec_grad}, where the effect is likely to be the
largest due to the faintness of the diffuse emission there, only
changed our results by an amount comparable to the errors (for
instance, the best fit temperature decreased by 0.02 keV).
We conclude that the difference in the shape of the spectra between
our true background and the blank sky background data is not large
enough to significantly affect our spectral fitting results, and
therefore do not include this component when doing spectral analyses
of diffuse emission.

A summary of the best fitting spectral models
is given in Table~\ref{tab:spectra} for different components of the
X-ray emission.
The first column gives the origin of the spectrum, the second column
lists the spectral model used, the third and fourth columns give the
temperature $T_s$ and abundances (if relevant) for the softer component
of the spectrum, the fifth column gives the power-law photon spectral
index $\Gamma$ or temperature $T_h$ of the harder component in the
spectrum, the sixth column gives the value of $\chi^2$ and the number of
degrees of freedom (dof), and the last column gives the number of net
counts (after background subtraction) in the spectrum.

\subsection{X-ray Spectrum of Resolved Sources} \label{sec:src_spec}

\begin{figure}
\vskip4.0truein
\includegraphics{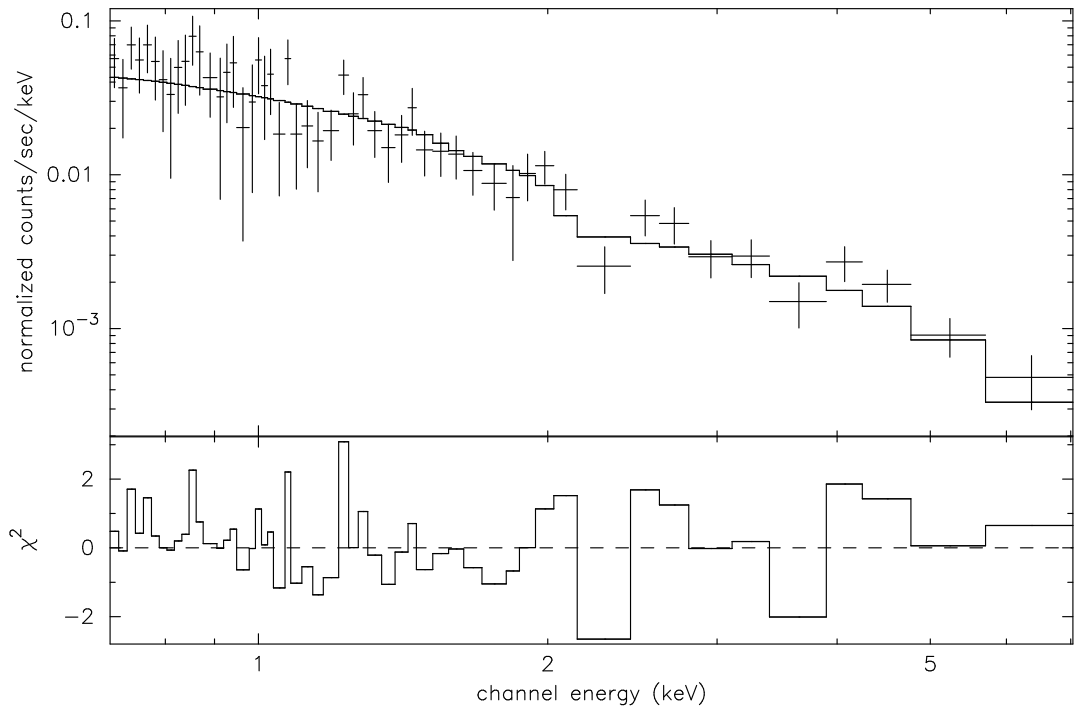}
\caption{X-ray spectrum of the sum of the sources within one $R_{\rm eff}$ of
NGC~4649, excluding Src.~1, fit with a model combining Galactic
absorption and a hard power-law.  The points with the error bars are
the data and the histogram shows the fitted model.
The lower panel shows the individual bins contributions to the chi-squared
of the fit.
\label{fig:src_spec}}
\end{figure}

Figure~\ref{fig:src_spec} shows the spectrum of the sum of the sources
on the S3 chip within one effective radius ($R_{\rm eff} = 82\arcsec$)
of the center of NGC~4649.
If the LMXBs are distributed like the stars then roughly 50\% of the
LMXBs will be contained within this radius, whereas we expect to find
$\la$1 background AGN in this region.  Choosing this radius has the
additional advantage that it facilitates comparisons of our results
with results from optical observations of NGC~4649.
For this source spectrum, we only
considered photon energies of 0.7--8.0 keV since the bins with
energies $\geq 8$ keV had very large errors.

We first considered models in which the spectrum of the sources was
represented by a single, hard component.
Initially, we tried a thermal bremsstrahlung model
(``bremss'' in Table~\ref{tab:spectra}) with a temperature $T_h$.
While this model gave a statistically significant fit, with a $\chi^2$ per
dof of $\chi^2_\nu = 0.97$, the temperature $T_h$ was poorly constrained.
As an alternative, we also fit the source spectrum with a power law
with a photon index of $\Gamma$.
This gave an acceptable fit to the spectrum which was slightly better
than the fit given by the thermal bremsstrahlung model, with a $\chi^2$ per
dof of $\chi^2_\nu = 0.83$.
We will adopt this as our best-fit model for the spectrum of the
source population.
We also tried a model for the sources with a power-law hard component
and a blackbody (``bbody'') soft component.
This gave a slightly better fit to the spectrum, but had very large
uncertainties in the power-law index and blackbody temperature.
Thus, we prefer the simpler power-law model.

\subsection{Diffuse X-ray Spectrum} \label{sec:diffuse_spec}

\begin{figure}
\vskip4.0truein
\includegraphics{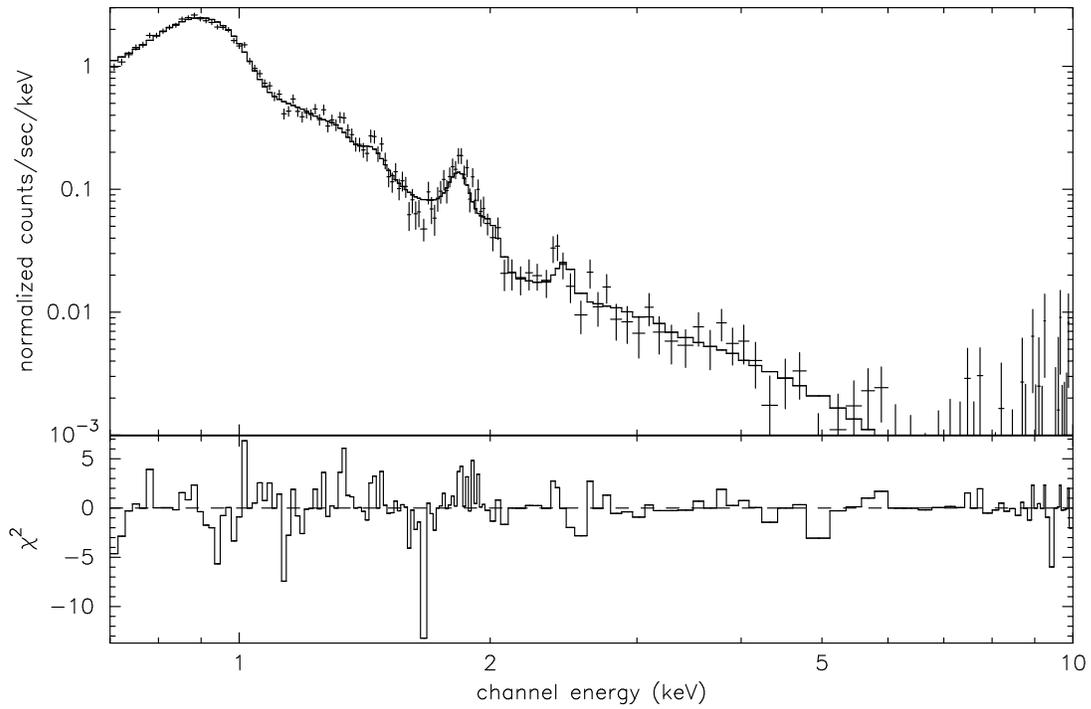}
\caption{X-ray spectrum of the diffuse emission within one $R_{\rm eff}$ of
NGC~4649, fit with a model combining Galactic absorption, a soft mekal
component, and a hard power-law component with the same spectral shape
as that of the resolved sources (Figure~\protect\ref{fig:src_spec}).
\label{fig:diffuse_spec}}
\end{figure}

We extracted the diffuse emission within one $R_{\rm eff}$ of
the center of NGC~4649 (Figure~\ref{fig:diffuse_spec}),
excluding all sources except Src.~1.
The X-ray color of the diffuse emission in this region is soft,
with hardness ratios of (H21,H31) $\approx (-0.381,-0.915)$.
The spectrum shows strong emission lines.
The most prominent are the helium-like
\ion{Si}{13} lines ($\sim$1.85 keV) and
\ion{S}{15} lines ($\sim$2.45 keV).
Also present in the spectra are
\ion{Mg}{12} lines and a blend of Fe L lines.
The strong lines indicate that the diffuse radiation is mainly
thermal emission from interstellar gas, as expected in this X-ray
bright elliptical galaxy.

Based on the line emitting spectrum, we first tried a single temperature
mekal model.
This did not provide an acceptable fit.
In particular, the residuals showed that there was a significant excess
of hard X-ray emission.
Since we would expect some hard X-ray emission from point sources below
our detection limit, we tried modeling this hard component with
a power-law spectrum.
This provided a much improved fit, and removed the excess hard emission.
The best-fit photon exponent was $\Gamma \approx 1.76$, which is very
close to the value found for the resolved sources;
fixing the photon exponent to the best fit value found in
\S~\ref{sec:src_spec},
$\Gamma \approx 1.78$, does not significantly
alter the fit so we assumed this value for consistency.
This best-fit mekal model, shown in
Figure~\ref{fig:diffuse_spec}, gave a gas temperature of $kT_s \approx 0.796$
and a heavy element abundance of about 0.58 solar.
We interpret this fit as representing soft
emission from hot diffuse gas combined with a hard power-law emission
from unresolved sources.

This model still has a $\chi^2_\nu$ which is larger than unity.
Examining the residuals, one finds that a substantial portion of the
$\chi^2$ originates in residuals at the positions of lines.
For example, the Si and S lines in the model spectrum are too weak,
and the Fe lines are a bit too strong.
This suggests that the ratios of the heavy elements in this galaxy
do not simply scale with the solar ratios.
To include wider abundance variations, we fit the soft component of
the spectrum with a vmekal model, in which each of the heavy element
abundances could vary independently.
We also fixed the normalization of the vmekal component to
the normalization given for the mekal component from the soft mekal
plus hard power-law model just discussed.  Allowing this normalization
to vary did somewhat improve the $\chi^2$ of the fit, but it gave
abundance values which were unreasonably high.
While the vmekal model improved the fit, most of the thirteen abundances
considered in this model were very
poorly constrained.
Examining the abundances from this fit suggested that the elements
could be divided into two groups:
the light elements (C through Ar) and the iron group elements (Ca, Fe,
and Ni).
This model, with the normalization allowed to vary, gave a
somewhat better fit to the spectrum, with a gas
temperature of $kT_s \approx 0.78$ keV and an abundance of
about 1.68 solar for the light elements and
1.12 solar for the iron group elements
(see also Table~\ref{tab:spectra}).
It may be surprising that these abundance values are both higher than
the abundance found above by the best-fit mekal model (one might have
expected the overall abundance to be an average of the abundances for
the light element and iron group elements).  This occurs because
separating the elements into two groups allows for more freedom in
determining the normalizations of each model component.  As a result
the vmekal normalization is decreased relative to the normalization of
the mekal model found above and the abundances must be increased to
match the observed line strengths.
Since this vmekal model still did not provide a good fit to the
spectrum of the diffuse emission, we tried adding a cooling flow
component (mkcflow) to the mekal plus power-law model described above,
with the high temperature and abundance tied to the temperature and
abundance values of the mekal component and the lower temperature frozen
at $kT_{\rm low} = 0.0808$ keV.  Adding the mkcflow component did not
improve the $\chi^2$ of the fit.  We found an upper limit on the mass
accretion rate of $\dot{M} < 0.1 \, M_{\odot}$ yr$^{-1}$.

\subsection{Radial Variation in the Spectrum of the Diffuse Component} \label{sec:radial}

The spectral properties of the diffuse emission as a function of
radius were examined to search for radial abundance and temperature
gradients in the gas.
The area within 120\arcsec\ of the galactic center
was broken up into 9 concentric annuli each containing
roughly 2000 net counts.
The outer annuli have more gross counts (but a higher fraction of
background), so that these spectra have more grouped bins and a larger
number of dofs.
Since the mekal and vmekal models provided roughly statistically
equivalent fits to the total diffuse emission (see \S~\ref{sec:diffuse_spec}),
we choose to fit each region with the simpler
mekal plus power-law spectral model with $\Gamma$ fixed at 1.78.
The results of the fits are given in Table~\ref{tab:spec_grad}.
Although the temperature of the diffuse
gas does appear to vary slightly as a function of radius, there is no
clear trend with temperature or abundance as a function of radius.
If one excludes the central circle
(where the fit is poor),
there is a stronger trend for
the temperature to increase slightly with increasing radius.

While most of the fits in Table~\ref{tab:spec_grad} are adequate,
the spectral fit to the central circle is not.
This suggests that the central region may require an additional component
or components to accurately model its spectrum.
Note that the central circle contains enhanced diffuse emission due
to a ``bar'' (\S~\ref{sec:central_bar}).
An examination of the residuals of the mekal plus power-law fit to the
central region revealed that the model was systematically fainter than
the data at energies $\la$ 0.9 keV.  We therefore tried adding a cool
component to the mekal plus power-law model to accommodate the data.
First we tried adding a second mekal component,
with the abundances of the two thermal components being equal, which
improved the fit.
Allowing the abundances
to vary independently did not significantly alter the fit.
The two temperatures in this model were $0.752^{+0.064}_{-0.047}$ and
$1.22^{+0.27}_{-0.16}$ keV, with
an abundance of $1.49^{+1.30}_{-0.74}$ solar.
The $\chi^2$ of the two-temperature fit
is 67.9 for 44 degrees of freedom, and while this is a significant
improvement over the single temperature model it still is not a
particularly good fit.
We therefore also tried a mekal plus power-law plus cooling flow
(mkcflow) model with the abundances and higher temperatures
of the mekal and mkcflow components set equal to one another and the
lower temperature set to $kT_{\rm low} = 0.0808$ keV.
While the $\chi^2$ of this fit was 78.0 for 45 degrees of freedom,
better than the
mekal plus power-law model, the $\chi^2$ of this fit was not as good as
that of the two mekal component model.  In addition, the abundance
was very poorly constrained.
This fit gave a temperature of $kT = 1.02^{+0.07}_{-0.06}$ keV, and
an upper limit on the mass accretion rate of
$\dot{M} < 0.054 \, M_{\odot}$ yr$^{-1}$.
We note that other authors have also found a preference for
two-temperature models over cooling flow and single-phase models in
the centers of elliptical galaxies (e.g., Buote 2002).

\section{Structure in the Diffuse Emission} \label{sec:diffuse_struct}

\subsection{Radial Features} \label{sec:radial_features}

\begin{figure}
\plotone{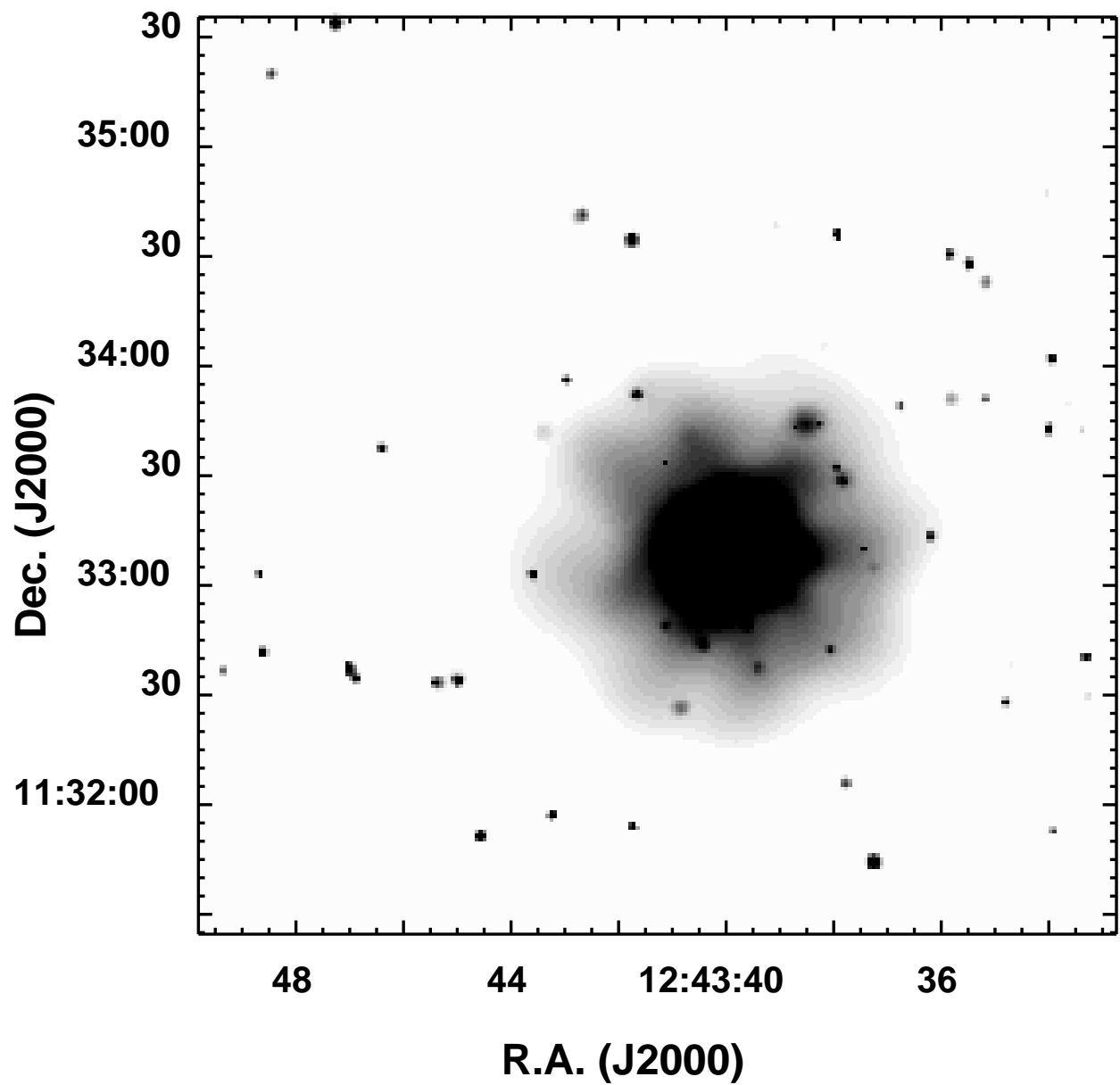}
\caption{Adaptively smoothed image of the central region
($4\farcm25 \times 4\farcm15$)
of NGC~4649
cleaned of background flares and
corrected for background and exposure.
Faint radial features can be seen extending out from the center of the
galaxy.
The greyscale is logarithmic and ranges from
$6.3 \times 10^{-6}$ to $3.6 \times 10^{-5}$
cnt pix$^{-1}$ s$^{-1}$.
\label{fig:soft_nuc}}
\end{figure}

\begin{figure}
\plotone{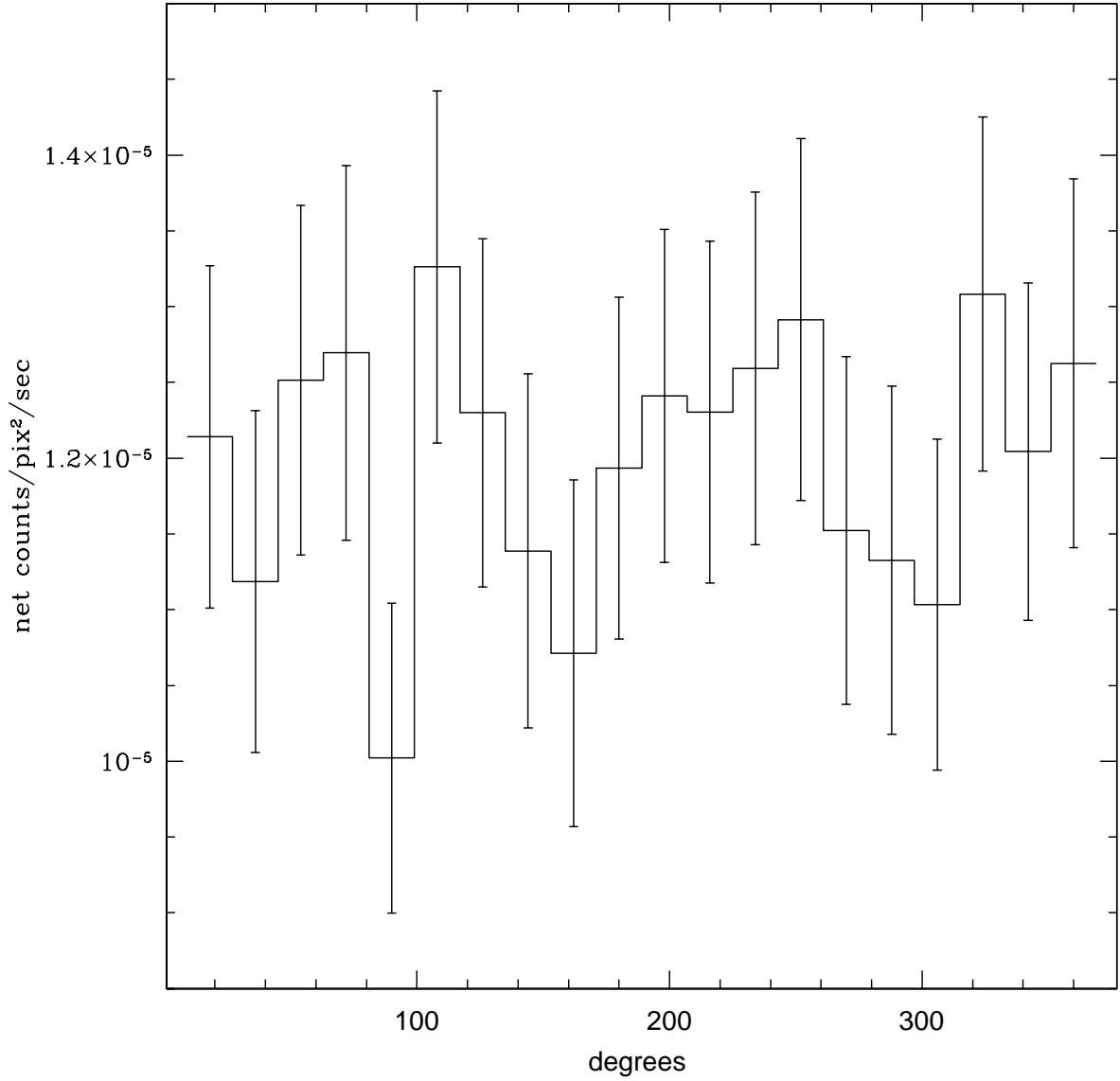}
\caption{Azimuthal plot of the net flux in 20 angular bins between
$21\arcsec$ and $53\arcsec$ from the center of NGC~4649.  These counts
are background subtracted but not corrected for exposure.  Angles are
measured from North to East.  Error bars are 90\% confidence intervals.
\label{fig:fingers}}
\end{figure}

The smoothed image in Figure~\ref{fig:xray_smo} shows some evidence for
structure in the diffuse emission near the core of NGC~4649.
Specifically, there appeared to be faint ``fingers'' extending out from
the core, between about 21\arcsec\ and 53\arcsec\ from the center of
the diffuse emission.
These radial features are shown more clearly in Figure~\ref{fig:soft_nuc},
which is an adaptively smoothed image of the center of NGC~4649.
These features are far too strong to be due to the support structure of
the telescope\footnote{See
\url{http://asc.harvard.edu/caldb/cxcpsflib.manual.ps}.}.
To check the statistical significance of these features, we defined a
series of regions containing the most obvious fingers.
We considered only the raw data, unsmoothed and uncorrected for exposure, to ensure that the fingers were not an
artifact of the smoothing process.
Detected point sources were removed from all of the regions.
We then compared the average surface brightness of these regions to that of
other regions in between the fingers at the same distance from the center of
NGC~4649.  For the regions containing the fingers, we found on average
$(2.06 \pm 0.06) \times 10^{-5}$ cnt s$^{-1}$ pix$^{-1}$, and
$(1.71 \pm 0.06) \times 10^{-5}$ cnt s$^{-1}$ pix$^{-1}$
for the regions without fingers
(1-$\sigma$ errors).
The excess of counts in the regions containing fingers was therefore
significant at the 4.1-$\sigma$ level.
As a further test of the
significance of these features, we examined the azimuthal brightness
profile in an annulus ranging from 21\arcsec\ to 53\arcsec\ centered on
the peak of the diffuse emission.  We then divided this annulus into 20 angular
bins of 18\arcdeg\ each and did a $\chi^2$ significance test comparing
the counts in each bin to the mean number of counts per bin
(an azimuthal plot of the flux in each bin is given in Figure~\ref{fig:fingers}).  We
found a reduced $\chi^2$ of
$\chi^2_\nu = 1.58$,
which suggests that
the azimuthal brightness profile is not well described by a constant.
It is therefore possible that the radial fingers we observed are real
features in the diffuse emission.  The hardness ratios of the fingers,
(H21,H31) $ = (-0.39,-0.90)$, and the hardness ratios of the regions
between the fingers (H21,H31) $ = (-0.41,-0.96)$ were both similar to
the hardness ratios of the total diffuse emission within one effective
radius (see \S~\ref{sec:diffuse_spec}).

Radial features have been produced in some numerical hydrodynamical
simulations of cooling flows in elliptical galaxies
(Kritsuk, B\"ohringer, \& M\"uller 1998, hereafter KBM).
KBM give results from two-dimensional axisymmetric hybrid inflow-outflow
models which show such features.
The physical extent of their features ($\la 7$ kpc) is about the same as
those we observe ($\la 5$ kpc).
Their model predicts an azimuthally-averaged
temperature profile which is depressed in the central region of the galaxy
and rises out to about 10 kpc (see their Fig.~11).
We find some evidence for a temperature profile which rises with radius,
excluding the central 4\arcsec\ of the
galaxy where the X-ray spectrum may be complicated by the presence of
the X-ray bar discussed in \S~\ref{sec:central_bar} (see
Table~\ref{tab:spec_grad}).
The simulations of KBM predict outflowing jets of hot gas, surrounded by
inflowing gas which is significantly cooler than the
outflowing gas at the same distance from the galactic center (see
their Fig.~7).
We therefore fit spectra accumulated from the sum of the finger regions
and from the regions in between the fingers, using the same mekal plus
power-law model used to fit the spectrum of the diffuse emission.
The abundances in these fits were fixed at 0.6 solar.  If the
abundances were allowed to vary the resulting values
were unreasonably high and had large errors.
The temperatures and $\chi^2$'s given by the fits were
not significantly affected by fixing the abundance values.
The results of these fits are given in Table~\ref{tab:spectra}.
While the bright fingers are slightly hotter than the regions
between the fingers, the 90\% confidence intervals overlap.
In the KBM model, one would expect larger temperature differences
of about a factor of two,  although projection effects might somewhat
reduce these differences.
More seriously, in the KBM model the bright regions would be considerably
cooler than the faint regions, whereas the bright regions are slightly
hotter in the observations.
Thus, while some form of convective instability may explain the fingers
in NGC~4649,
only the morphology is reproduced by the KBM model.

\subsection{Central Bar} \label{sec:central_bar}

\begin{figure}
\plotone{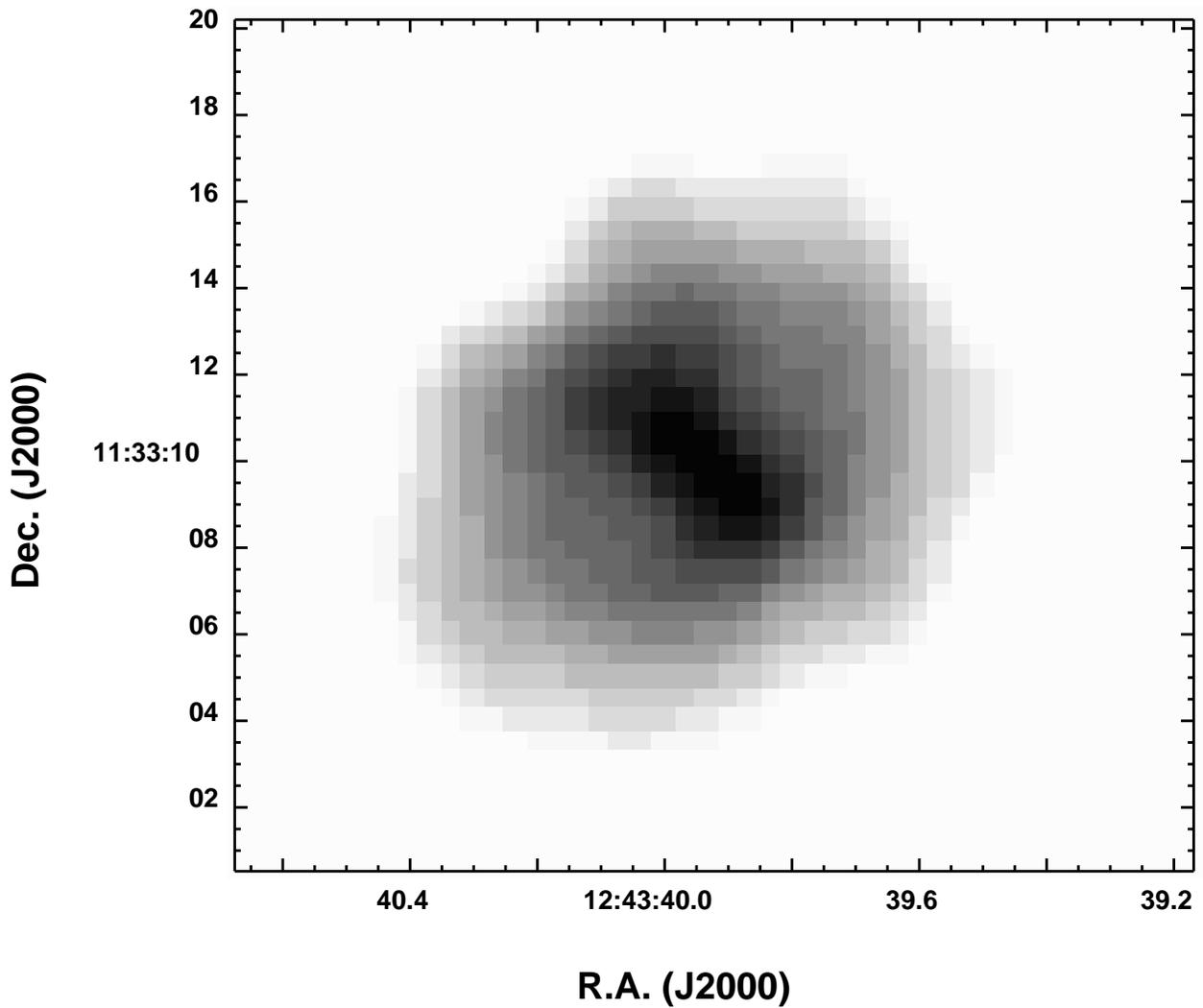}
\caption{Magnified view of the very center
($22\arcsec \times 20\arcsec$)
of the adaptively smoothed image shown in Figure~\protect\ref{fig:soft_nuc},
showing a roughly 5\arcsec\ long bar running from the SW to the NE at the
center of the galaxy.
The greyscale is logarithmic and ranges from $1.8 \times 10^{-4}$
to $6.9 \times 10^{-4}$ cnt pix$^{-1}$ s$^{-1}$.
\label{fig:bar}}
\end{figure}

Figure~\ref{fig:bar} shows an even smaller region at the nucleus of
NGC~4649 from the same image shown in (Figure~\ref{fig:soft_nuc}).  There
appears to be a small bar, approximately 5\arcsec\  long, located at the
central peak in the diffuse emission.
To assess the significance of the elongation of this feature, we determined
the net counts in an elliptical region corresponding to this bar and
compared it to the net counts in an identical region perpendicular to
and crossing the center of the bar.
The counts in the area where these two regions overlapped were excluded.
We found $283 \pm 18$ net counts in the bar region
and $223 \pm 16$ net counts in the perpendicular off-bar region
(1-$\sigma$ errors).
While this is not a large difference, it is marginally statistically
significant (a 2.5-$\sigma$ detection).
The bar appears to be oriented roughly perpendicular to the X-ray isophotes at
larger radii, which go from SW to NE.  It also appears to be
somewhat asymmetric, with the SW end being somewhat brighter than the
NE end (if we identify the brightest point in the bar as its center).
The bar is included in the region given for Src.~1 as detected by the
source detection algorithm we used (see \S~\ref{sec:src_detect}),
although the region of Src.~1 is not centered
on nor rotationally aligned with the (much smaller) bar.  Radio observations of the
center of NGC~4649 with the Very Large Array show an extended object,
$7\arcsec \times 3\arcsec$ in size oriented 31\arcdeg\ East of North
(Condon et al.\ 1991), making it coincident with the X-ray bar we
observe.
The radio source has an integrated flux density of 18 mJy at 4.86 GHz.
A recent unrelated study of the {\it Chandra} data has detected a soft
(0.2--0.6 keV) central source that the authors interpret as being a
quiescent supermassive black hole (Soldatenkov, Vikhlinin, \&
Pavlinsky 2003).  Although no exact position for this
source is given it is said to be at the peak of the optical emission,
and therefore presumably near to or contained within the X-ray bar we observe.
Unfortunately, there are too few counts from the bar to
determine its spectral properties.  We can, however, determine the
X-ray colors of the bar.
We find that the X-ray color is medium-soft, with
(H21,H31) $\approx (-0.164 \pm 0.033,-0.896 \pm 0.019$,
1-$\sigma$ errors).
This is somewhat harder than the color of the diffuse emission within one
effective radius, (H21,H31) $\approx (-0.381 \pm 0.007,-0.915 \pm 0.006)$).
It is also slightly harder than the color of the perpendicular off-bar
region, although they are consistent within the errors.
However, the X-ray emission is probably not hard enough to indicate a
strong shock in the region of the bar.

There are several possible explanations for this bar.
First, the bar may be an emission feature, or due to excess absorption
perpendicular to the bar.
If the bar were due to excess absorption perpendicular to the bar,
the off-bar region would have a particularly hard spectrum, whereas the
hardness ratios indicate that the bar is slightly harder than
the off-bar region.
In addition, when doing spectral fits to the central region of NGC~4649,
allowing the local absorption to vary did not improve the fit.

If the bar is an emission feature, it might be a rotating disk or
torus viewed edge-on, perhaps produced by the cooling of rotating X-ray
gas
(e.g., Brighenti \& Mathews 2000).
However, if this were the case one would expect the bar to be cooler than
the surrounding gas, which is not true here.
Alternatively, the bar
could be a shock associated with the expansion of a central radio
source, as appears to have occurred in NGC~4636
(Jones et al.\ 2002).
Although the X-ray colors of the bar are not
hard enough to indicate a strong shock, they are somewhat harder than
the surrounding diffuse emission.  Furthermore, the bar appears to be
coincident with the extended radio source at the center of the galaxy,
although the radio emission is relatively weak.  Although we did not
detect a central X-ray source which could be associated with an AGN,
the recent detection by Soldatenkov et al.\ (2003), which they interpret
as a quiescent supermassive black hole, could be responsible for the
radio structure and the X-ray bar.  However, it should be noted that
the interpretation of the central source as an LMXB cannot be
completely ruled out due to its low luminosity ($L_X = 6.0 \times
10^{37}$ ergs s$^{-1}$ in the 0.2--0.6 keV band).  Still, the
interpretation of the bar as resulting from compression
from the radio source is probably the most consistent explanation given
the existing data.

\section{Conclusion} \label{sec:conclusion}

We have presented results from a {\it Chandra} observation of the X-ray
bright elliptical galaxy NGC~4649.  Both bright diffuse emission and
point sources are detected, with the bright diffuse emission dominating the
overall emission.

A total of 165 discrete sources were detected.
Fourteen of these sources had possible faint
optical counterparts in the DSS image of the same field.
Of these fourteen, three had positions listed in the USNO-A2.0 catalog.
In a region to the NW of the center of NGC~4649 we find roughly 20 X-ray
sources whose positions match those of globular clusters seen in HST
images.  In this region, roughly 45\% of the X-ray sources are
identified with globular clusters.
Some of the sources may be associated with the companion Sc galaxy
NGC~4647.
However, we do not detect X-rays from the Type I supernova
SN1979a in NGC~4649.
We do detect the serendipitous {\it ROSAT} source RX J1244.1+1134 as
a blend of three sources.

The luminosity function of the resolved sources is well described by a
broken power-law, with a break luminosity which is somewhat larger than
the Eddington luminosity for a 1.4 $M_{\odot}$ neutron star.
This break luminosity is slightly larger than previously determined values
for other early-type galaxies
(Sarazin et al.\ 2000;
Blanton et al.\ 2001), although it is consistent with previous
measurements within the errors.

Emission from diffuse gas swamps emission from discrete sources near
the center of NGC~4649.
No point sources are detected within 12\arcsec\ of
the center of the galaxy.
We find no conclusive evidence for a central AGN,
although another study of the {\it Chandra} data has found evidence
for a central quiescent supermassive black hole.
We give an upper limit of
$L_X$ (0.3--10 keV) $\le 3.3 \times 10^{38}$ erg s$^{-1}$ for the
X-ray luminosity of any AGN.

The composite X-ray spectrum of the resolved sources within 1 $R_{\rm
eff}$ is best described by a power-law with a photon spectral index of
$\Gamma \approx 1.78$.
The spectrum of the diffuse emission is best fit by a mekal plus power-law
model, with a gas temperature $kT \approx 0.80$ keV,
an abundance of roughly 58\% of solar,
and $\Gamma$ frozen at 1.78.
This argues that the diffuse emission is
a combination of emission from diffuse gas and unresolved LMXBs.
There is some evidence that the abundances of iron group elements
are lower than those of lighter elements when compared to the solar
ratios.

There is also evidence for
radial features in the diffuse emission extending out from the
center of NGC~4649, and for a central bar.
Two-dimensional hydrodynamic simulations of cooling flows in
elliptical galaxies by Kritsuk et al.\ (1998) predict such radial
features, which result from convective motions with
relatively hot outflowing gas separated by regions of cooler
inflowing gas.
However, only the morphology of the observed fingers match the
specific predictions of the KBM model.

The central bar is roughly perpendicular to the X-ray and optical isophotes
at larger radii, is parallel to the extension in the weak central
radio source, and may contain a recently detected quiescent
supermassive black hole.
We suggest that the X-ray bar may be produced by an interaction between the
central radio source and the surrounding X-ray gas.

\acknowledgements
We are very grateful to Arunav Kundu for providing us with his
unpublished list of globular clusters in NGC~4649, and for
several very helpful conversations.
Support for this work was provided by the National Aeronautics and Space
Administration primarily through $Chandra$ Award Number
GO0-1141X,
but also through
GO1-2078X,
both
issued by the $Chandra$ X-ray Observatory Center, which is operated by the
Smithsonian Astrophysical Observatory for and on behalf of NASA under
contract
NAS8-39073.

%
%
\begin{table}[p]
\tiny
\caption{\hfil Discrete X-ray Sources \label{tab:src} \hfil}
\begin{center}
\begin{tabular}{lcccrrrrccl}
\tableline
\tableline
Src.&
Name&
R.A.&
Dec.&
\multicolumn{1}{c}{$d$}&
\multicolumn{1}{c}{Count Rate}&
\multicolumn{1}{c}{SNR}&
\multicolumn{1}{c}{$L_X$}&
H21&
H31&
\\
No.&
&
(h:m:s)&
($\arcdeg$:$\arcmin$:$\arcsec$)&
\multicolumn{1}{c}{($\arcsec$)}&
\multicolumn{1}{c}{($10^{-4}$ s$^{-1}$)}&
&
&
&
&
Notes \\
(1)&
(2)&
(3)&
(4)&
\multicolumn{1}{c}{(5)}&
\multicolumn{1}{c}{(6)}&
\multicolumn{1}{c}{(7)}&
\multicolumn{1}{c}{(8)}&
(9)&
(10)&
(11)
\\
\tableline
   1&CXOU J124340.0+113311&12:43:40.01&11:33:11.8&  1.63&641.07$\pm$13.87   &46.21&         1441.35& \nodata               &   \nodata             &a\\
   2&CXOU J124339.2+113317&12:43:39.25&11:33:17.4& 13.48& 33.52$\pm$\phn3.46& 9.70&   \phn\phn75.36&$-0.56^{+0.17}_{-0.13}$&$-0.88^{+0.15}_{-0.07}$&b\\
   3&CXOU J124340.2+113324&12:43:40.22&11:33:24.7& 14.79& 20.43$\pm$\phn2.90& 7.04&   \phn\phn45.94&$-0.26^{+0.32}_{-0.27}$&$-0.40^{+0.28}_{-0.22}$&b,f\\
   4&CXOU J124340.0+113252&12:43:40.02&11:32:52.0& 18.19& 20.80$\pm$\phn2.92& 7.13&   \phn\phn46.77&$-0.38^{+0.38}_{-0.28}$&$-0.33^{+0.28}_{-0.23}$&b\\
   5&CXOU J124340.9+113325&12:43:40.99&11:33:25.2& 20.65& 13.11$\pm$\phn2.50& 5.23&   \phn\phn29.47&$-0.15^{+0.33}_{-0.30}$&$-0.63^{+0.41}_{-0.22}$&b\\
   6&CXOU J124338.6+113304&12:43:38.64&11:33:04.3& 21.19& 21.49$\pm$\phn3.01& 7.14&   \phn\phn48.31&$+0.20^{+0.25}_{-0.28}$&$-0.49^{+0.35}_{-0.24}$&b,f\\
   7&CXOU J124341.3+113320&12:43:41.34&11:33:20.5& 21.91& 10.52$\pm$\phn2.12& 4.97&   \phn\phn23.66&$-0.39^{+0.31}_{-0.24}$&$-0.89^{+0.65}_{-0.10}$&b\\
   8&CXOU J124339.6+113248&12:43:39.61&11:32:48.4& 22.62& 16.46$\pm$\phn2.55& 6.46&   \phn\phn37.01&$-0.42^{+0.51}_{-0.34}$&$+0.01^{+0.32}_{-0.33}$&b\\
   9&CXOU J124338.3+113305&12:43:38.38&11:33:05.4& 24.64& 15.94$\pm$\phn2.57& 6.20&   \phn\phn35.85&$-0.14^{+0.25}_{-0.24}$&$-0.35^{+0.25}_{-0.21}$&b\\
  10&CXOU J124341.1+113248&12:43:41.11&11:32:48.8& 26.68& 10.79$\pm$\phn2.10& 5.15&   \phn\phn24.26&$-0.30^{+0.41}_{-0.32}$&$-0.81^{+0.75}_{-0.17}$&b\\
  11&CXOU J124340.4+113243&12:43:40.45&11:32:43.3& 27.61& 38.27$\pm$\phn3.75&10.21&   \phn\phn86.05&$-0.08^{+0.15}_{-0.14}$&$-0.60^{+0.18}_{-0.13}$&b\\
  12&CXOU J124341.1+113333&12:43:41.14&11:33:33.3& 28.33&  9.94$\pm$\phn1.98& 5.02&   \phn\phn22.36&$+0.16^{+0.47}_{-0.56}$&$-0.15^{+0.63}_{-0.52}$&b\\
  13&CXOU J124342.0+113321&12:43:42.00&11:33:21.8& 31.26&  6.55$\pm$\phn1.67& 3.93&   \phn\phn14.73&$+0.38^{+0.49}_{-0.88}$&$-0.60^{+1.48}_{-0.39}$&b\\
  14&CXOU J124339.4+113238&12:43:39.40&11:32:38.0& 33.47& 11.86$\pm$\phn2.27& 5.22&   \phn\phn26.66&$-0.08^{+0.28}_{-0.27}$&$-0.44^{+0.34}_{-0.25}$&b\\
  15&CXOU J124338.8+113241&12:43:38.89&11:32:41.1& 33.53&  8.55$\pm$\phn1.82& 4.69&   \phn\phn19.23&$-0.65^{+0.37}_{-0.20}$&$-0.51^{+0.27}_{-0.19}$&b\\
  16&CXOU J124337.6+113309&12:43:37.68&11:33:09.9& 34.46&  4.06$\pm$\phn1.28& 3.16&\phn\phn\phn9.13&$-1.00^{+2.57}_{-0.00}$&$-0.20^{+1.07}_{-0.74}$&b\\
  17&CXOU J124339.4+113235&12:43:39.40&11:32:35.0& 36.36&  7.21$\pm$\phn1.70& 4.25&   \phn\phn16.21&$-0.21^{+0.28}_{-0.25}$&$-0.51^{+0.35}_{-0.24}$&b\\
  18&CXOU J124337.8+113328&12:43:37.83&11:33:28.4& 37.04& 20.77$\pm$\phn2.80& 7.41&   \phn\phn46.71&$+0.01^{+0.31}_{-0.31}$&$-0.15^{+0.33}_{-0.30}$&b,f\\
  19&CXOU J124337.9+113332&12:43:37.94&11:33:32.1& 37.67&  9.43$\pm$\phn2.00& 4.73&   \phn\phn21.21&$-0.04^{+0.29}_{-0.28}$&$-1.00^{+0.38}_{-0.00}$&b\\
  20&CXOU J124338.7+113342&12:43:38.71&11:33:42.6& 37.73& 15.20$\pm$\phn2.36& 6.45&   \phn\phn34.17&$-0.01^{+0.32}_{-0.32}$&$-0.14^{+0.34}_{-0.31}$&b,f\\
  21&CXOU J124339.6+113347&12:43:39.68&11:33:47.6& 37.75&  6.38$\pm$\phn1.69& 3.78&   \phn\phn14.34&$-0.08^{+0.49}_{-0.45}$&$-0.27^{+0.58}_{-0.43}$&b\\
  22&CXOU J124342.2+113329&12:43:42.25&11:33:29.8& 38.13&  5.15$\pm$\phn1.43& 3.61&   \phn\phn11.59&$+0.08^{+0.29}_{-0.30}$&$-0.50^{+0.46}_{-0.29}$&b,f\\
  23&CXOU J124337.4+113309&12:43:37.43&11:33:09.2& 38.15&  9.29$\pm$\phn1.85& 5.02&   \phn\phn20.88&$+1.00^{+0.00}_{-1.20}$&$+1.00^{+0.00}_{-1.16}$&b,f\\
  24&CXOU J124341.1+113234&12:43:41.11&11:32:34.6& 39.00&  5.51$\pm$\phn1.52& 3.63&   \phn\phn12.40&$-1.00^{+0.69}_{-0.00}$&$-0.23^{+0.67}_{-0.50}$&b\\
  25&CXOU J124338.0+113242&12:43:38.06&11:32:42.2& 40.21& 16.21$\pm$\phn2.36& 6.88&   \phn\phn36.45&$+0.27^{+0.24}_{-0.28}$&$-0.02^{+0.31}_{-0.30}$&b\\
  26&CXOU J124342.2+113335&12:43:42.21&11:33:35.0& 40.58&  6.69$\pm$\phn1.66& 4.03&   \phn\phn15.05&$+0.28^{+0.44}_{-0.60}$&$+0.11^{+0.52}_{-0.59}$&b\\
  27&CXOU J124338.6+113345&12:43:38.60&11:33:45.0& 40.62&  7.68$\pm$\phn1.75& 4.39&   \phn\phn17.26&$-0.07^{+0.23}_{-0.22}$&$-0.51^{+0.30}_{-0.21}$&b\\
  28&CXOU J124337.2+113304&12:43:37.25&11:33:04.3& 41.20&  7.39$\pm$\phn1.72& 4.29&   \phn\phn16.61&$-0.38^{+0.38}_{-0.28}$&$-0.03^{+0.27}_{-0.27}$&b,f\\
  29&CXOU J124338.2+113343&12:43:38.27&11:33:43.9& 42.44& 11.72$\pm$\phn2.10& 5.59&   \phn\phn26.36&$+0.12^{+0.20}_{-0.21}$&$-0.63^{+0.35}_{-0.20}$&b,f\\
  30&CXOU J124340.8+113226&12:43:40.85&11:32:26.3& 45.53&  5.46$\pm$\phn1.54& 3.54&   \phn\phn12.27&$-0.47^{+0.53}_{-0.32}$&$-0.50^{+0.52}_{-0.31}$&\\
  31&CXOU J124341.6+113351&12:43:41.65&11:33:51.9& 48.07& 18.10$\pm$\phn2.49& 7.26&   \phn\phn40.70&$+0.07^{+0.20}_{-0.21}$&$-0.49^{+0.30}_{-0.21}$&f\\
  32&CXOU J124342.6+113340&12:43:42.62&11:33:40.7& 48.84&  7.53$\pm$\phn1.77& 4.26&   \phn\phn16.94&$-0.45^{+0.44}_{-0.29}$&$-0.48^{+0.44}_{-0.29}$&f\\
  33&CXOU J124336.9+113329&12:43:36.96&11:33:29.9& 49.17&  6.16$\pm$\phn1.60& 3.85&   \phn\phn13.84&$-0.36^{+0.49}_{-0.35}$&$-0.79^{+1.12}_{-0.20}$&f\\
  34&CXOU J124343.3+113319&12:43:43.34&11:33:19.2& 49.55&  4.84$\pm$\phn1.39& 3.48&   \phn\phn10.89&$+0.73^{+0.27}_{-1.67}$&$-0.50^{+1.50}_{-0.50}$&\\
  35&CXOU J124343.0+113245&12:43:43.03&11:32:45.1& 50.79&  5.87$\pm$\phn1.51& 3.89&   \phn\phn13.20&$-0.90^{+1.81}_{-0.10}$&$-0.42^{+0.48}_{-0.32}$&\\
  36&CXOU J124342.3+113231&12:43:42.31&11:32:31.7& 51.08&  3.57$\pm$\phn1.17& 3.05&\phn\phn\phn8.02&$-0.18^{+0.62}_{-0.51}$&$-0.54^{+0.94}_{-0.39}$&\\
  37&CXOU J124340.0+113218&12:43:40.02&11:32:18.5& 51.69&  5.51$\pm$\phn1.47& 3.74&   \phn\phn12.39&$-0.47^{+0.96}_{-0.45}$&$-0.38^{+0.69}_{-0.43}$&\\
  38&CXOU J124341.3+113221&12:43:41.39&11:32:21.9& 52.29&  5.19$\pm$\phn1.45& 3.58&   \phn\phn11.67&$-0.56^{+0.78}_{-0.35}$&$-0.56^{+0.75}_{-0.34}$&\\
  39&CXOU J124343.5+113302&12:43:43.59&11:33:02.8& 52.91& 24.02$\pm$\phn2.80& 8.59&   \phn\phn54.00&$-0.30^{+0.18}_{-0.16}$&$-0.55^{+0.20}_{-0.15}$&\\
  40&CXOU J124338.0+113355&12:43:38.08&11:33:55.0& 53.15&  5.23$\pm$\phn1.45& 3.62&   \phn\phn11.76&$+0.16^{+0.54}_{-0.66}$&$+0.07^{+0.60}_{-0.66}$&\\
  41&CXOU J124336.6+113251&12:43:36.61&11:32:51.0& 53.73&  5.92$\pm$\phn1.48& 3.99&   \phn\phn13.31&$+1.00^{+0.00}_{-0.72}$&$+1.00^{+0.00}_{-2.42}$&\\
  42&CXOU J124336.1+113313&12:43:36.19&11:33:13.1& 56.43& 11.38$\pm$\phn2.01& 5.67&   \phn\phn25.58&$+0.16^{+0.27}_{-0.30}$&$-0.33^{+0.43}_{-0.32}$&f\\
  43&CXOU J124341.6+113218&12:43:41.69&11:32:18.5& 57.19&  5.10$\pm$\phn1.43& 3.58&   \phn\phn11.46&$+0.24^{+0.42}_{-0.54}$&$-0.89^{+1.89}_{-0.11}$&\\
  44&CXOU J124343.8+113255&12:43:43.82&11:32:55.7& 57.62&  4.91$\pm$\phn1.43& 3.44&   \phn\phn11.05&$-0.41^{+0.45}_{-0.31}$&$-1.00^{+0.36}_{-0.00}$&\\
  45&CXOU J124343.3+113341&12:43:43.36&11:33:41.6& 58.22&  6.42$\pm$\phn1.52& 4.23&   \phn\phn14.43&$+0.11^{+0.33}_{-0.35}$&$-0.36^{+0.50}_{-0.35}$&f\\
  46&CXOU J124338.2+113405&12:43:38.20&11:34:05.4& 61.38&  6.96$\pm$\phn1.63& 4.28&   \phn\phn15.66&$+0.59^{+0.28}_{-0.55}$&$+0.27^{+0.47}_{-0.66}$&f\\
  47&CXOU J124336.7+113348&12:43:36.75&11:33:48.6& 61.58&  5.88$\pm$\phn1.51& 3.91&   \phn\phn13.22&$-0.66^{+0.74}_{-0.27}$&$-0.48^{+0.55}_{-0.33}$&f\\
  48&CXOU J124342.9+113356&12:43:42.96&11:33:56.1& 63.00&  8.74$\pm$\phn1.72& 5.09&   \phn\phn19.66&$+0.10^{+0.35}_{-0.38}$&$+0.04^{+0.37}_{-0.38}$&f\\
  49&CXOU J124340.3+113413&12:43:40.30&11:34:13.5& 63.44&  4.89$\pm$\phn1.36& 3.60&   \phn\phn11.00&$+0.49^{+0.37}_{-0.69}$&$-0.67^{+1.64}_{-0.33}$&\\
  50&CXOU J124344.6+113315&12:43:44.61&11:33:15.8& 67.62&  4.77$\pm$\phn1.41& 3.39&   \phn\phn10.72&$+0.25^{+0.40}_{-0.51}$&$-0.26^{+0.72}_{-0.51}$&\\
  51&CXOU J124344.8+113254&12:43:44.82&11:32:54.6& 72.17&  6.82$\pm$\phn1.53& 4.46&   \phn\phn15.33&$+0.29^{+0.28}_{-0.35}$&$-0.26^{+0.55}_{-0.42}$&\\
  52&CXOU J124337.7+113205&12:43:37.78&11:32:05.3& 72.79&  5.34$\pm$\phn1.38& 3.88&   \phn\phn12.00&$+0.92^{+0.08}_{-1.76}$&$+0.69^{+0.31}_{-1.66}$&\\
  53&CXOU J124335.8+113350&12:43:35.80&11:33:50.7& 74.13& 10.12$\pm$\phn1.94& 5.22&   \phn\phn22.75&$+0.03^{+0.35}_{-0.36}$&$+0.02^{+0.36}_{-0.36}$&f\\
  54&CXOU J124345.1+113316&12:43:45.10&11:33:16.2& 74.83&  4.48$\pm$\phn1.32& 3.40&   \phn\phn10.08&$+0.27^{+0.39}_{-0.50}$&$-0.03^{+0.57}_{-0.55}$&\\
  55&CXOU J124341.2+113157&12:43:41.24&11:31:57.4& 74.95&  4.01$\pm$\phn1.21& 3.31&\phn\phn\phn9.02&$+0.08^{+0.38}_{-0.41}$&$-1.00^{+1.05}_{-0.00}$&\\
  56&CXOU J124341.7+113154&12:43:41.73&11:31:54.0& 80.20& 13.79$\pm$\phn2.19& 6.30&   \phn\phn31.00&$-0.45^{+0.23}_{-0.19}$&$-0.98^{+1.97}_{-0.02}$&\\
  57&CXOU J124341.5+113428&12:43:41.52&11:34:28.5& 81.34&  3.89$\pm$\phn1.16& 3.34&\phn\phn\phn8.74&$+0.10^{+0.51}_{-0.57}$&$-0.47^{+1.07}_{-0.46}$&f\\
  58&CXOU J124335.2+113350&12:43:35.20&11:33:50.5& 81.56&  9.33$\pm$\phn1.76& 5.30&   \phn\phn20.97&$-0.13^{+0.35}_{-0.32}$&$-0.25^{+0.40}_{-0.33}$&\\
  59&CXOU J124344.9+113233&12:43:44.99&11:32:33.7& 81.58& 43.23$\pm$\phn3.60&12.01&   \phn\phn97.19&$-0.10^{+0.11}_{-0.11}$&$-0.30^{+0.12}_{-0.11}$&\\
  60&CXOU J124345.5+113319&12:43:45.56&11:33:19.8& 81.91&  3.42$\pm$\phn1.12& 3.04&\phn\phn\phn7.68&$+0.74^{+0.24}_{-1.24}$&$-0.14^{+1.12}_{-0.84}$&\\
\tableline
\end{tabular}
\end{center}
\end{table}

%
%
\begin{table}[p]
\tiny
\begin{center}
\begin{tabular}{lcccrrrrccl}
\multicolumn{8}{c}{\normalsize Table \protect\ref{tab:src}---$Continued$}\\
\tableline
\tableline
Src.&
Name&
R.A.&
Dec.&
\multicolumn{1}{c}{$d$}&
\multicolumn{1}{c}{Count Rate}&
\multicolumn{1}{c}{SNR}&
\multicolumn{1}{c}{$L_X$}&
H21&
H31&
\\
No.&
&
(h:m:s)&
($\arcdeg$:$\arcmin$:$\arcsec$)&
\multicolumn{1}{c}{($\arcsec$)}&
\multicolumn{1}{c}{($10^{-4}$ s$^{-1}$)}&
&
&
&
&
Notes \\
(1)&
(2)&
(3)&
(4)&
\multicolumn{1}{c}{(5)}&
\multicolumn{1}{c}{(6)}&
\multicolumn{1}{c}{(7)}&
\multicolumn{1}{c}{(8)}&
(9)&
(10)&
(11)
\\
\tableline
  61&CXOU J124340.9+113431&12:43:40.94&11:34:31.0& 81.92&  3.85$\pm$\phn1.16& 3.31&\phn\phn\phn8.65&$+0.37^{+0.40}_{-0.60}$&$-0.84^{+1.83}_{-0.16}$&\\
  62&CXOU J124334.3+113310&12:43:34.39&11:33:10.6& 82.81&  4.41$\pm$\phn1.21& 3.65&\phn\phn\phn9.92&$+1.00^{+0.00}_{-0.97}$&$+1.00^{+0.00}_{-1.21}$&f\\
  63&CXOU J124334.6+113237&12:43:34.68&11:32:37.3& 85.16&  9.51$\pm$\phn1.74& 5.47&   \phn\phn21.38&$-0.42^{+0.34}_{-0.25}$&$-0.40^{+0.33}_{-0.25}$&f\\
  64&CXOU J124345.5+113340&12:43:45.50&11:33:40.9& 86.12&  5.70$\pm$\phn1.41& 4.04&   \phn\phn12.81&$-0.10^{+0.35}_{-0.32}$&$-1.00^{+0.32}_{-0.00}$&\\
  65&CXOU J124343.5+113420&12:43:43.52&11:34:20.2& 86.83&  3.75$\pm$\phn1.18& 3.19&\phn\phn\phn8.43&$-0.89^{+1.51}_{-0.11}$&$-0.59^{+0.59}_{-0.28}$&d\\
  66&CXOU J124345.3+113232&12:43:45.37&11:32:32.6& 87.09& 11.59$\pm$\phn1.96& 5.92&   \phn\phn26.05&$+0.03^{+0.28}_{-0.29}$&$+0.06^{+0.27}_{-0.28}$&\\
  67&CXOU J124343.2+113156&12:43:43.23&11:31:56.8& 87.21& 14.73$\pm$\phn2.13& 6.93&   \phn\phn33.12&$+0.22^{+0.20}_{-0.22}$&$-0.43^{+0.38}_{-0.27}$&\\
  68&CXOU J124334.9+113356&12:43:34.98&11:33:56.4& 87.36&  4.87$\pm$\phn1.28& 3.81&   \phn\phn10.94&$-0.21^{+0.46}_{-0.38}$&$-0.37^{+0.53}_{-0.37}$&\\
  69&CXOU J124341.7+113434&12:43:41.75&11:34:34.1& 87.66& 38.18$\pm$\phn3.38&11.30&   \phn\phn85.84&$-0.13^{+0.10}_{-0.10}$&$-0.64^{+0.12}_{-0.10}$&\\
  70&CXOU J124334.7+113227&12:43:34.79&11:32:27.6& 87.94& 19.65$\pm$\phn2.44& 8.05&   \phn\phn44.18&$+0.18^{+0.16}_{-0.17}$&$-0.44^{+0.26}_{-0.20}$&f\\
  71&CXOU J124335.3+113405&12:43:35.32&11:34:05.0& 88.23&  4.21$\pm$\phn1.25& 3.36&\phn\phn\phn9.47&$+0.58^{+0.40}_{-1.37}$&$+0.28^{+0.70}_{-1.20}$&\\
  72&CXOU J124339.0+113438&12:43:39.06&11:34:38.5& 89.44&  6.34$\pm$\phn1.52& 4.16&   \phn\phn14.26&$-0.23^{+0.43}_{-0.36}$&$-0.31^{+0.44}_{-0.34}$&\\
  73&CXOU J124333.9+113319&12:43:33.91&11:33:19.1& 90.30&  3.35$\pm$\phn1.11& 3.02&\phn\phn\phn7.52&$+0.77^{+0.23}_{-1.52}$&$+0.25^{+0.74}_{-1.21}$&\\
  74&CXOU J124337.9+113435&12:43:37.94&11:34:35.5& 90.65& 16.57$\pm$\phn2.26& 7.32&   \phn\phn37.25&$-0.02^{+0.21}_{-0.21}$&$-0.10^{+0.22}_{-0.21}$&\\
  75&CXOU J124334.0+113342&12:43:34.03&11:33:42.2& 93.73& 41.29$\pm$\phn3.53&11.70&   \phn\phn92.84&$-0.05^{+0.11}_{-0.11}$&$-0.37^{+0.13}_{-0.12}$&\\
  76&CXOU J124337.2+113143&12:43:37.27&11:31:43.9& 95.31& 67.09$\pm$\phn4.41&15.20&      \phn150.83&$-0.27^{+0.08}_{-0.08}$&$-0.54^{+0.08}_{-0.07}$&h\\
  77&CXOU J124346.4+113337&12:43:46.40&11:33:37.4& 97.56&  9.99$\pm$\phn1.89& 5.30&   \phn\phn22.47&$+0.82^{+0.17}_{-0.97}$&$+0.85^{+0.14}_{-0.90}$&\\
  78&CXOU J124342.8+113439&12:43:42.81&11:34:39.8& 98.52&  6.69$\pm$\phn1.54& 4.36&   \phn\phn15.04&$+0.23^{+0.39}_{-0.48}$&$+0.08^{+0.48}_{-0.52}$&\\
  79&CXOU J124342.6+113441&12:43:42.65&11:34:41.3& 98.94& 11.47$\pm$\phn1.93& 5.94& \phn\phn25.79&$+0.17^{+0.28}_{-0.31}$&$+0.21^{+0.27}_{-0.30}$&\\
  80&CXOU J124335.8+113430&12:43:35.82&11:34:30.1&101.02& 10.88$\pm$\phn1.88& 5.78&   \phn\phn24.47&$+0.06^{+0.22}_{-0.22}$&$-0.52^{+0.39}_{-0.25}$&f\\
  81&CXOU J124333.6+113348&12:43:33.66&11:33:48.9&101.23&  9.14$\pm$\phn1.76& 5.20&   \phn\phn20.55&$-0.16^{+0.24}_{-0.22}$&$-0.97^{+1.96}_{-0.03}$&\\
  82&CXOU J124335.1+113422&12:43:35.17&11:34:22.4&101.51&  7.57$\pm$\phn1.62& 4.66&   \phn\phn17.01&$-0.21^{+0.36}_{-0.31}$&$-0.30^{+0.42}_{-0.33}$&f,g\\
  83&CXOU J124335.4+113427&12:43:35.48&11:34:27.5&102.17& 24.30$\pm$\phn2.71& 8.98&   \phn\phn54.64&$+0.03^{+0.15}_{-0.15}$&$-0.30^{+0.19}_{-0.17}$&f,g\\
  84&CXOU J124333.3+113342&12:43:33.38&11:33:42.2&102.77&  8.62$\pm$\phn1.66& 5.18&   \phn\phn19.38&$+0.59^{+0.27}_{-0.52}$&$+0.48^{+0.34}_{-0.59}$&\\
  85&CXOU J124333.9+113401&12:43:33.94&11:34:01.4&103.05& 17.74$\pm$\phn2.37& 7.50&   \phn\phn39.88&$+0.08^{+0.17}_{-0.18}$&$-0.53^{+0.29}_{-0.20}$&\\
  86&CXOU J124333.3+113240&12:43:33.30&11:32:40.1&103.31& 31.18$\pm$\phn3.04&10.26&   \phn\phn70.10&$-0.02^{+0.13}_{-0.13}$&$-0.11^{+0.14}_{-0.14}$&\\
  87&CXOU J124344.5+113151&12:43:44.55&11:31:51.0&103.41& 21.69$\pm$\phn2.57& 8.44&   \phn\phn48.77&$+0.08^{+0.18}_{-0.19}$&$+0.12^{+0.18}_{-0.18}$&\\
  88&CXOU J124342.1+113130&12:43:42.19&11:31:30.9&104.26&  6.01$\pm$\phn1.41& 4.26&   \phn\phn13.52&$-0.40^{+0.51}_{-0.35}$&$-0.10^{+0.38}_{-0.35}$&\\
  89&CXOU J124333.0+113246&12:43:33.09&11:32:46.9&104.54&  5.33$\pm$\phn1.32& 4.04&   \phn\phn11.99&$+0.07^{+0.33}_{-0.35}$&$-0.24^{+0.46}_{-0.38}$&\\
  90&CXOU J124346.8+113234&12:43:46.89&11:32:34.2&107.12& 20.95$\pm$\phn2.53& 8.29&   \phn\phn47.09&$-0.10^{+0.17}_{-0.17}$&$-0.29^{+0.19}_{-0.17}$&\\
  91&CXOU J124333.2+113229&12:43:33.27&11:32:29.5&107.29&  6.04$\pm$\phn1.40& 4.31&   \phn\phn13.58&$-0.10^{+0.31}_{-0.29}$&$-0.32^{+0.41}_{-0.32}$&\\
  92&CXOU J124347.0+113237&12:43:47.02&11:32:37.0&108.03& 44.02$\pm$\phn3.60&12.22&   \phn\phn98.98&$-0.01^{+0.11}_{-0.11}$&$-0.18^{+0.12}_{-0.12}$&\\
  93&CXOU J124333.4+113212&12:43:33.41&11:32:12.8&112.89&  3.57$\pm$\phn1.12& 3.19&\phn\phn\phn8.02&$-0.33^{+0.58}_{-0.40}$&$-0.47^{+0.67}_{-0.37}$&\\
  94&CXOU J124341.5+113117&12:43:41.55&11:31:17.3&115.09&  4.53$\pm$\phn1.24& 3.66&   \phn\phn10.18&$-0.33^{+0.70}_{-0.46}$&$-0.12^{+0.55}_{-0.48}$&\\
  95&CXOU J124335.6+113447&12:43:35.63&11:34:47.3&116.63&  4.20$\pm$\phn1.20& 3.52&\phn\phn\phn9.45&$-0.74^{+0.76}_{-0.22}$&$-0.94^{+1.94}_{-0.06}$&d,g\\
  96&CXOU J124333.9+113152&12:43:33.92&11:31:52.6&118.62&  8.15$\pm$\phn1.61& 5.06&   \phn\phn18.32&$-0.37^{+0.32}_{-0.25}$&$-0.28^{+0.32}_{-0.27}$&\\
  97&CXOU J124342.7+113118&12:43:42.70&11:31:18.2&118.69&  3.35$\pm$\phn1.07& 3.12&\phn\phn\phn7.54&$+0.53^{+0.41}_{-1.00}$&$+0.29^{+0.62}_{-1.00}$&\\
  98&CXOU J124334.9+113443&12:43:34.95&11:34:43.0&119.06&  8.70$\pm$\phn1.66& 5.23&   \phn\phn19.55&$-0.32^{+0.33}_{-0.27}$&$-0.34^{+0.34}_{-0.27}$&g,h\\
  99&CXOU J124347.5+113222&12:43:47.54&11:32:22.0&120.50&  5.36$\pm$\phn1.33& 4.04&   \phn\phn12.05&$+0.00^{+0.43}_{-0.43}$&$-0.25^{+0.58}_{-0.44}$&\\
 100&CXOU J124344.3+113125&12:43:44.30&11:31:25.8&121.84&  5.50$\pm$\phn1.38& 3.97&   \phn\phn12.37&$-0.30^{+0.48}_{-0.36}$&$-0.73^{+1.10}_{-0.25}$&\\
 101&CXOU J124336.0+113119&12:43:36.04&11:31:19.8&124.96& 10.60$\pm$\phn1.83& 5.80&   \phn\phn23.84&$-0.24^{+0.28}_{-0.24}$&$-0.17^{+0.28}_{-0.25}$&d\\
 102&CXOU J124343.4+113115&12:43:43.43&11:31:15.6&125.04& 10.03$\pm$\phn1.76& 5.70&   \phn\phn22.56&$+0.02^{+0.26}_{-0.26}$&$-0.29^{+0.37}_{-0.30}$&\\
 103&CXOU J124331.9+113350&12:43:31.95&11:33:50.0&125.17&  3.69$\pm$\phn1.12& 3.30&\phn\phn\phn8.29&$-0.45^{+0.45}_{-0.30}$&$-0.39^{+0.43}_{-0.31}$&\\
 104&CXOU J124348.6+113302&12:43:48.68&11:33:02.9&127.41& 13.52$\pm$\phn2.02& 6.68&   \phn\phn30.40&$+0.08^{+0.21}_{-0.22}$&$-0.22^{+0.28}_{-0.25}$&\\
 105&CXOU J124331.3+113303&12:43:31.34&11:33:03.2&127.83& 13.38$\pm$\phn2.02& 6.61&   \phn\phn30.08&$+0.03^{+0.23}_{-0.23}$&$+0.09^{+0.22}_{-0.23}$&\\
 106&CXOU J124340.9+113517&12:43:40.96&11:35:17.4&127.95&  4.70$\pm$\phn1.29& 3.64&   \phn\phn10.56&$-0.14^{+0.43}_{-0.38}$&$-0.11^{+0.42}_{-0.39}$&\\
 107&CXOU J124348.6+113241&12:43:48.61&11:32:41.3&129.44& 16.47$\pm$\phn2.23& 7.38&   \phn\phn37.03&$-0.13^{+0.18}_{-0.18}$&$-0.62^{+0.26}_{-0.17}$&d,f\\
 108&CXOU J124334.0+113446&12:43:34.07&11:34:46.7&130.28&  5.22$\pm$\phn1.36& 3.83&   \phn\phn11.73&$-0.40^{+0.60}_{-0.38}$&$-0.05^{+0.44}_{-0.42}$&g\\
 109&CXOU J124347.8+113206&12:43:47.83&11:32:06.5&131.20&  5.11$\pm$\phn1.30& 3.93&   \phn\phn11.50&$+0.12^{+0.38}_{-0.42}$&$-0.24^{+0.57}_{-0.44}$&d\\
 110&CXOU J124332.0+113418&12:43:32.07&11:34:18.8&135.55& 27.51$\pm$\phn2.87& 9.57&   \phn\phn61.85&$+0.61^{+0.12}_{-0.15}$&$+0.50^{+0.15}_{-0.19}$&g\\
 111&CXOU J124330.9+113338&12:43:30.97&11:33:38.0&135.94&  4.90$\pm$\phn1.24& 3.94&   \phn\phn11.01&$-0.35^{+0.47}_{-0.34}$&$-0.35^{+0.51}_{-0.36}$&\\
 112&CXOU J124349.3+113237&12:43:49.35&11:32:37.0&141.01&  8.54$\pm$\phn1.64& 5.21&   \phn\phn19.19&$+0.14^{+0.25}_{-0.27}$&$-0.33^{+0.44}_{-0.33}$&\\
 113&CXOU J124332.0+113146&12:43:32.02&11:31:46.9&144.14& 31.66$\pm$\phn4.16& 7.62&   \phn\phn71.19&$-0.36^{+0.18}_{-0.15}$&$-0.60^{+0.20}_{-0.15}$&c\\
 114&CXOU J124340.6+113045&12:43:40.69&11:30:45.9&144.62&  6.19$\pm$\phn1.39& 4.44&   \phn\phn13.92&$-0.32^{+0.41}_{-0.32}$&$-0.63^{+0.80}_{-0.30}$&\\
 115&CXOU J124344.4+113056&12:43:44.45&11:30:56.0&149.12&  3.08$\pm$\phn1.01& 3.06&\phn\phn\phn6.93&$-0.33^{+0.53}_{-0.38}$&$-0.57^{+0.88}_{-0.35}$&\\
 116&CXOU J124329.8+113318&12:43:29.85&11:33:18.5&149.76& 23.26$\pm$\phn2.61& 8.90&   \phn\phn52.31&$+0.03^{+0.17}_{-0.17}$&$+0.09^{+0.17}_{-0.17}$&\\
 117&CXOU J124344.7+113056&12:43:44.79&11:30:56.9&150.57&  6.84$\pm$\phn1.53& 4.46&   \phn\phn15.38&$-0.49^{+0.40}_{-0.27}$&$-0.41^{+0.45}_{-0.31}$&d\\
 118&CXOU J124334.8+113520&12:43:34.83&11:35:20.8&151.28&  3.56$\pm$\phn1.07& 3.31&\phn\phn\phn7.99&$+0.43^{+0.37}_{-0.60}$&$+0.05^{+0.62}_{-0.66}$&g\\
 119&CXOU J124332.2+113450&12:43:32.27&11:34:50.5&151.82&  7.98$\pm$\phn1.62& 4.93&   \phn\phn17.95&$+0.07^{+0.36}_{-0.38}$&$-0.18^{+0.46}_{-0.39}$&g\\
 120&CXOU J124349.8+113218&12:43:49.85&11:32:18.3&153.43&  6.86$\pm$\phn1.47& 4.68&   \phn\phn15.43&$+0.40^{+0.27}_{-0.37}$&$+0.15^{+0.39}_{-0.44}$&\\
\tableline
\end{tabular}
\end{center}
\end{table}

%
%
\begin{table}[p]
\tiny
\begin{center}
\begin{tabular}{lcccrrrrccl}
\multicolumn{8}{c}{\normalsize Table \protect\ref{tab:src}---$Continued$}\cr
\tableline
\tableline
Src.&
Name&
R.A.&
Dec.&
\multicolumn{1}{c}{$d$}&
\multicolumn{1}{c}{Count Rate}&
\multicolumn{1}{c}{SNR}&
\multicolumn{1}{c}{$L_X$}&
H21&
H31&
\\
No.&
&
(h:m:s)&
($\arcdeg$:$\arcmin$:$\arcsec$)&
\multicolumn{1}{c}{($\arcsec$)}&
\multicolumn{1}{c}{($10^{-4}$ s$^{-1}$)}&
&
&
&
&
Notes \\
(1)&
(2)&
(3)&
(4)&
\multicolumn{1}{c}{(5)}&
\multicolumn{1}{c}{(6)}&
\multicolumn{1}{c}{(7)}&
\multicolumn{1}{c}{(8)}&
(9)&
(10)&
(11)
\\
\tableline
 121&CXOU J124341.3+113037&12:43:41.30&11:30:37.6&153.74&  3.38$\pm$\phn1.04& 3.24&\phn\phn\phn7.61&$+0.20^{+0.62}_{-0.83}$&$+0.54^{+0.36}_{-0.78}$&\\
 122&CXOU J124339.9+113036&12:43:39.95&11:30:36.2&153.99&  4.11$\pm$\phn1.15& 3.57&\phn\phn\phn9.24&$+0.42^{+0.36}_{-0.57}$&$-0.16^{+0.79}_{-0.63}$&\\
 123&CXOU J124343.2+113537&12:43:43.29&11:35:37.0&154.45&  3.84$\pm$\phn1.14& 3.38&\phn\phn\phn8.64&$-0.08^{+0.43}_{-0.40}$&$-0.35^{+0.60}_{-0.41}$&\\
 124&CXOU J124334.7+113527&12:43:34.79&11:35:27.6&157.48&  3.41$\pm$\phn1.08& 3.16&\phn\phn\phn7.66&$-0.71^{+0.51}_{-0.21}$&$-1.00^{+0.00}_{-0.00}$&g\\
 125&CXOU J124340.3+113547&12:43:40.36&11:35:47.9&157.79&  5.46$\pm$\phn1.47& 3.71&   \phn\phn12.28&$-0.20^{+0.40}_{-0.34}$&$-0.69^{+0.90}_{-0.27}$&\\
 126&CXOU J124345.3+113531&12:43:45.34&11:35:31.7&161.64&  4.40$\pm$\phn1.30& 3.38&\phn\phn\phn9.90&$-0.27^{+0.39}_{-0.32}$&$-1.00^{+0.56}_{-0.00}$&\\
 127&CXOU J124348.4+113110&12:43:48.46&11:31:10.9&172.04&  4.98$\pm$\phn1.24& 4.00&   \phn\phn11.19&$-0.56^{+0.69}_{-0.32}$&$-0.14^{+0.42}_{-0.37}$&\\
 128&CXOU J124329.8+113440&12:43:29.89&11:34:40.6&174.23&  3.94$\pm$\phn1.12& 3.52&\phn\phn\phn8.87&$+0.82^{+0.16}_{-0.86}$&$+0.38^{+0.56}_{-1.12}$&g\\
 129&CXOU J124335.6+113026&12:43:35.62&11:30:26.8&175.75& 29.01$\pm$\phn6.08& 4.84&   \phn\phn65.22&$-0.18^{+0.29}_{-0.26}$&$-1.00^{+0.20}_{-0.00}$&c\\
 130&CXOU J124351.6+113357&12:43:51.68&11:33:57.0&177.56& 14.97$\pm$\phn2.17& 6.90&   \phn\phn33.66&$-0.23^{+0.20}_{-0.18}$&$-0.36^{+0.25}_{-0.21}$&\\
 131&CXOU J124347.2+113533&12:43:47.26&11:35:33.6&178.53& 16.99$\pm$\phn2.31& 7.36&   \phn\phn38.20&$+0.13^{+0.17}_{-0.18}$&$-0.43^{+0.30}_{-0.23}$&\\
 132&CXOU J124348.4+113519&12:43:48.45&11:35:19.8&179.24& 12.18$\pm$\phn2.01& 6.05&   \phn\phn27.39&$-0.31^{+0.24}_{-0.20}$&$-0.55^{+0.32}_{-0.21}$&\\
 133&CXOU J124341.2+113010&12:43:41.28&11:30:10.7&180.43&  3.11$\pm$\phn1.01& 3.07&\phn\phn\phn7.00&$+0.13^{+0.46}_{-0.52}$&$-0.53^{+1.23}_{-0.44}$&\\
 134&CXOU J124327.9+113402&12:43:27.96&11:34:02.3&184.80&  7.13$\pm$\phn1.47& 4.84&   \phn\phn16.04&$-0.19^{+0.27}_{-0.25}$&$-0.24^{+0.30}_{-0.26}$&\\
 135&CXOU J124342.9+113009&12:43:42.93&11:30:09.9&185.28& 10.33$\pm$\phn1.76& 5.88&   \phn\phn23.23&$-0.30^{+0.25}_{-0.22}$&$-0.56^{+0.34}_{-0.22}$&\\
 136&CXOU J124336.4+113009&12:43:36.49&11:30:09.3&188.20&359.49$\pm$   22.32&16.36&      \phn808.27&$-0.23^{+0.07}_{-0.07}$&$-0.64^{+0.07}_{-0.06}$&c,d,e,f\\
 137&CXOU J124352.9+113312&12:43:52.97&11:33:12.2&190.25&  8.63$\pm$\phn1.71& 5.05&   \phn\phn19.40&$-0.10^{+0.23}_{-0.22}$&$-1.00^{+0.04}_{-0.00}$&\\
 138&CXOU J124335.5+113609&12:43:35.55&11:36:09.2&190.71& 13.83$\pm$\phn2.15& 6.44&   \phn\phn31.10&$+0.21^{+0.20}_{-0.21}$&$-0.07^{+0.27}_{-0.26}$&\\
 139&CXOU J124353.1+113301&12:43:53.17&11:33:01.2&193.39&  4.12$\pm$\phn1.24& 3.32&\phn\phn\phn9.27&$-0.66^{+0.68}_{-0.26}$&$-0.38^{+0.54}_{-0.36}$&\\
 140&CXOU J124344.2+113004&12:43:44.29&11:30:04.6&195.89& 12.07$\pm$\phn1.89& 6.38&   \phn\phn27.14&$+0.02^{+0.19}_{-0.19}$&$-0.08^{+0.21}_{-0.20}$&\\
 141&CXOU J124328.2+113443&12:43:28.25&11:34:43.9&196.78&  7.37$\pm$\phn1.51& 4.87&   \phn\phn16.57&$-0.25^{+0.34}_{-0.28}$&$-0.68^{+0.75}_{-0.26}$&g\\
 142&CXOU J124347.0+113022&12:43:47.03&11:30:22.3&196.94&  5.61$\pm$\phn1.35& 4.17&   \phn\phn12.62&$-0.09^{+0.34}_{-0.32}$&$-0.36^{+0.46}_{-0.33}$&\\
 143&CXOU J124326.6+113321&12:43:26.63&11:33:21.5&197.18&  9.46$\pm$\phn2.97& 3.19&   \phn\phn21.26&$-0.14^{+0.51}_{-0.45}$&$-0.19^{+0.55}_{-0.45}$&c\\
 144&CXOU J124352.0+113443&12:43:52.00&11:34:43.6&199.24&  4.48$\pm$\phn1.27& 3.54&   \phn\phn10.08&$-0.07^{+0.40}_{-0.38}$&$-0.92^{+1.92}_{-0.08}$&\\
 145&CXOU J124346.3+113011&12:43:46.32&11:30:11.6&201.13&  6.83$\pm$\phn1.46& 4.67&   \phn\phn15.35&$-0.39^{+0.33}_{-0.26}$&$-0.67^{+0.69}_{-0.25}$&\\
 146&CXOU J124330.5+113537&12:43:30.59&11:35:37.9&202.59&  3.78$\pm$\phn1.22& 3.10&\phn\phn\phn8.51&$+0.50^{+0.38}_{-0.78}$&$+0.22^{+0.59}_{-0.82}$&g\\
 147&CXOU J124354.5+113242&12:43:54.51&11:32:42.0&214.74& 12.77$\pm$\phn2.04& 6.27&   \phn\phn28.72&$-0.45^{+0.23}_{-0.18}$&$-0.64^{+0.33}_{-0.19}$&\\
 148&CXOU J124354.9+113307&12:43:54.99&11:33:07.8&219.94& 43.95$\pm$\phn3.64&12.08&   \phn\phn98.82&$-0.02^{+0.10}_{-0.10}$&$-0.61^{+0.14}_{-0.11}$&\\
 149&CXOU J124337.0+113646&12:43:37.06&11:36:46.4&220.56& 23.74$\pm$\phn2.95& 8.05&   \phn\phn53.38&$-0.08^{+0.15}_{-0.14}$&$-0.82^{+0.31}_{-0.12}$&\\
 150&CXOU J124350.5+113558&12:43:50.53&11:35:58.7&228.53&  5.78$\pm$\phn1.56& 3.71&   \phn\phn12.99&$-0.60^{+1.45}_{-0.39}$&$+0.23^{+0.52}_{-0.69}$&\\
 151&CXOU J124326.7+113514&12:43:26.78&11:35:14.9&231.16& 13.97$\pm$\phn2.14& 6.51&   \phn\phn31.40&$+0.12^{+0.21}_{-0.22}$&$-0.23^{+0.29}_{-0.25}$&g\\
 152&CXOU J124326.4+113521&12:43:26.40&11:35:21.7&239.55&  4.46$\pm$\phn1.27& 3.52&   \phn\phn10.04&$+0.11^{+0.42}_{-0.46}$&$-0.14^{+0.58}_{-0.50}$&g\\
 153&CXOU J124357.3+113301&12:43:57.33&11:33:01.7&254.46& 11.93$\pm$\phn2.00& 5.95&   \phn\phn26.83&$-0.01^{+0.21}_{-0.21}$&$-0.51^{+0.39}_{-0.25}$&\\
 154&CXOU J124341.9+113728&12:43:41.93&11:37:28.9&260.22&  6.01$\pm$\phn1.47& 4.08&   \phn\phn13.51&$+0.06^{+0.30}_{-0.32}$&$-0.23^{+0.47}_{-0.38}$&d\\
 155&CXOU J124342.1+113743&12:43:42.17&11:37:43.9&275.52&  4.18$\pm$\phn1.25& 3.34&\phn\phn\phn9.39&$-0.76^{+1.18}_{-0.22}$&$-0.05^{+0.42}_{-0.40}$&\\
 156&CXOU J124357.4+113450&12:43:57.48&11:34:50.9&275.57&  6.76$\pm$\phn1.63& 4.14&   \phn\phn15.20&$-0.25^{+0.34}_{-0.29}$&$-0.29^{+0.44}_{-0.34}$&\\
 157&CXOU J124347.7+113740&12:43:47.71&11:37:40.2&292.68& 10.99$\pm$\phn2.14& 5.13&   \phn\phn24.71&$+0.08^{+0.25}_{-0.27}$&$-0.21^{+0.41}_{-0.35}$&d,f\\
 158&CXOU J124345.0+113803&12:43:45.07&11:38:03.1&302.15&  8.32$\pm$\phn2.01& 4.14&   \phn\phn18.70&$+0.17^{+0.28}_{-0.31}$&$-0.93^{+1.93}_{-0.07}$&\\
 159&CXOU J124400.5+113230&12:44:00.57&11:32:30.0&304.60&  8.60$\pm$\phn1.93& 4.45&   \phn\phn19.34&$+0.05^{+0.38}_{-0.40}$&$+0.00^{+0.46}_{-0.46}$&\\
 160&CXOU J124356.9+113629&12:43:56.99&11:36:29.3&319.05&  6.10$\pm$\phn1.78& 3.42&   \phn\phn13.72&$-0.11^{+0.46}_{-0.42}$&$-0.27^{+0.64}_{-0.47}$&d,f\\
 161&CXOU J124349.6+113809&12:43:49.60&11:38:09.9&331.09& 13.57$\pm$\phn2.48& 5.47&   \phn\phn30.52&$-0.08^{+0.28}_{-0.26}$&$-0.30^{+0.47}_{-0.36}$&d\\
 162&CXOU J124347.9+113826&12:43:47.91&11:38:26.2&336.58&  7.41$\pm$\phn1.95& 3.80&   \phn\phn16.67&$-0.14^{+0.32}_{-0.29}$&$-0.79^{+1.48}_{-0.21}$&d,e,f\\
 163&CXOU J124355.6+113745&12:43:55.65&11:37:45.7&358.64&  5.29$\pm$\phn1.67& 3.18&   \phn\phn11.90&$+0.59^{+0.29}_{-0.63}$&$+0.06^{+0.73}_{-0.80}$&\\
 164&CXOU J124406.5+113420&12:44:06.53&11:34:20.2&395.75& 19.60$\pm$\phn3.14& 6.29&   \phn\phn44.06&$-0.19^{+0.21}_{-0.19}$&$-0.40^{+0.31}_{-0.24}$&d\\
 165&CXOU J124408.9+113333&12:44:08.95&11:33:33.2&425.71&200.41$\pm$\phn8.80&23.07&      \phn450.59&$-0.18^{+0.05}_{-0.05}$&$-0.49^{+0.06}_{-0.05}$&d,e\\
\tableline
\end{tabular}
\end{center}
\tablecomments{The units for $L_X$ are $10^{37}$ ergs s$^{-1}$ in the
0.3--10 keV band.}
\tablenotetext{a}{\scriptsize Src.~1 is is extended, and appears to be a
combination of a diffuse structure with one or more point sources.}
\tablenotetext{b}{\scriptsize Positions and count rates of sources
near the center of NGC~4649 are uncertain due to the bright diffuse emission
and/or confusion with nearby sources.}
\tablenotetext{c}{\scriptsize Source is at the edge of the S3 detector, and
flux is uncertain due to large exposure correction.}
\tablenotetext{d}{\scriptsize Possible faint optical counterpart.}
\tablenotetext{e}{\scriptsize Possible USNO-A2.0 optical counterpart.}
\tablenotetext{f}{\scriptsize Globular cluster is possible optical counterpart.}
\tablenotetext{g}{\scriptsize May be associated with the companion galaxy NGC~4647.}
\tablenotetext{h}{\scriptsize Source may be variable.}
\end{table}

\clearpage

\begin{deluxetable}{lcccccccc}
\tabletypesize{\small}
\tablewidth{6.75truein}
\tablecaption{Spectral Fits \label{tab:spectra}}
\tablehead{
\colhead{Origin}&
\colhead{Model}&
\colhead{$kT_s$}&
\colhead{Abund.\tablenotemark{a}}&
\colhead{$\Gamma$ or $k T_h$}&
\colhead{$\chi^2$/dof}&
\colhead{Net Cts.}\\
\colhead{}&
\colhead{}&
\colhead{(keV)}&
\colhead{(solar)}&
\colhead{(keV)}&
\colhead{}&
\colhead{}
}
\startdata
Sources&bremss&                         &                                             &5.81$^{+4.62}_{-2.02}$&49.5/51&804\\
Sources&powerlaw&                         &                                             &1.78$^{+0.19}_{-0.18}$&42.4/51  &804\\
Sources&powerlaw$+$bbody&1.51$^{+0.89}_{-0.49}$   &                                             &2.53$^{+0.62}_{-0.33}$&37.3/49  &804\\
Diffuse&mekal&0.805$^{+0.007}_{-0.007}$&0.33$^{+0.03}_{-0.03}$&&275.5/154&17187\\
Diffuse&mekal$+$powerlaw&0.796$^{+0.007}_{-0.007}$&0.58$^{+0.98}_{-0.15}$                       &1.76$^{+0.58}_{-0.80}$&203.1/152&17187\\
Diffuse&mekal$+$powerlaw&0.796$^{+0.007}_{-0.007}$&0.58$^{+0.15}_{-0.10}$                       &(1.78)                &203.2/153&17187\\
Diffuse&vmekal$+$powerlaw&0.784$^{+0.008}_{-0.008}$&1.68$^{+1.17}_{-0.77}$/1.12$^{+0.62}_{-0.28}$&(1.78)               &185.1/152&17187\\
Fingers&mekal$+$powerlaw&0.802$^{+0.032}_{-0.032}$&(0.6) &(1.78)                &35.0/33&1019\\
Off Fingers&mekal$+$powerlaw&0.756$^{+0.035}_{-0.041}$&(0.6) &(1.78)                &27.9/25&663\\
\enddata
\tablenotetext{a}{For the vmekal model, the first entry in this column gives
the abundance of the light elements C -- Ar, and the second entry is
for the iron group elements Ca, Fe, and Ni.}
\end{deluxetable}

\clearpage

\begin{deluxetable}{ccccc}
\tablewidth{3.75truein}
\tablecaption{Radial Variation in the Diffuse Spectrum \label{tab:spec_grad}}
\tablehead{
\colhead{Radii}&
\colhead{$kT$}&
\colhead{Abund.}&
\colhead{$\chi^2$/dof}&
\colhead{Net Cts.}\\
\colhead{(\arcsec)}&
\colhead{(keV)}&
\colhead{(solar)}&
\colhead{}&
\colhead{}
}
\startdata
0 - 4  &0.917$^{+0.034}_{-0.030}$&0.47$^{+0.14}_{-0.18}$&89.4/46&1897\\
4 - 7  &0.783$^{+0.019}_{-0.018}$&0.74$^{+0.19}_{-0.16}$&56.8/47&2258\\
7 - 11 &0.757$^{+0.021}_{-0.018}$&0.69$^{+0.13}_{-0.25}$&42.1/45&2380\\
11 - 17&0.740$^{+0.018}_{-0.017}$&0.76$^{+0.98}_{-0.13}$&56.2/46&2318\\
17 - 26&0.771$^{+0.010}_{-0.019}$&0.45$^{+0.18}_{-0.15}$&41.7/46&2185\\
26 - 38&0.802$^{+0.025}_{-0.021}$&0.57$^{+0.37}_{-0.12}$&46.3/44&1930\\
38 - 54&0.824$^{+0.022}_{-0.022}$&0.97$^{+10.4}_{-0.19}$&59.7/55&1856\\
54 - 80&0.848$^{+0.023}_{-0.022}$&0.69$^{+0.31}_{-0.14}$&96.0/74&2291\\
80 - 120&0.908$^{+0.026}_{-0.024}$&0.43$^{+0.12}_{-0.16}$&90.7/109&2469\\
\enddata
\end{deluxetable}

\clearpage

\clearpage

\clearpage

\clearpage

\clearpage

\end{document}